\documentclass{jfm}
\usepackage{amsmath}
\usepackage{amsfonts}
\usepackage{amssymb}
\usepackage{color}

\definecolor{grey0}{gray}{0.8}
\definecolor{grey1}{gray}{0.75}
\definecolor{grey2}{gray}{0.5}
\definecolor{grey3}{gray}{0.35}
\definecolor{grey4}{gray}{0.35}
\definecolor{blue1}{rgb}{0.88,0.88,1}
\definecolor{blue2}{rgb}{0.83,0.83,0.95}
\definecolor{blue3}{rgb}{0.8,0.8,0.9}
\definecolor{blue4}{rgb}{0.73,0.73,0.85}
\definecolor{blue5}{rgb}{0.6,0.6,0.8}
\definecolor{blue6}{rgb}{0.35,0.35,0.6}
\definecolor{blue7}{rgb}{0.2,0.2,0.45}
\definecolor{blue8}{rgb}{0.2,0.2,0.3}

\definecolor{blue9}{rgb}{0.4,0.6,1.}
\definecolor{blue10}{rgb}{0.1,0.1,0.9}
\definecolor{blue11}{rgb}{0.,0.,0.5}
\definecolor{red1}{rgb}{1,0.6,0.4}
\definecolor{red2}{rgb}{0.9,0.,0.}
\definecolor{red3}{rgb}{0.5,0.,0.}
\definecolor{green1}{rgb}{0.7,0.95,0.7}
\definecolor{green2}{rgb}{0.3,0.8,0.3}
\definecolor{green3}{rgb}{0.,0.5,0.}

\begin{document}

\newtheorem{lemma}{Lemma}
\newtheorem{corollary}{Corollary}

\shorttitle{Slope influence in turbulent bedload transport} 
\shortauthor{R. Maurin et al} 

\title{Revisiting slope influence in turbulent bedload transport: consequences for vertical flow structure and transport rate scaling}

\author
 {
Raphael Maurin\aff{1}
  \corresp{\email{raphael.maurin@imft.fr}}
  Julien Chauchat\aff{2,3}
  \and 
  Philippe Frey\aff{4}
  }

\affiliation
{
\aff{1}
Institut de M\'ecanique des Fluides de Toulouse, IMFT, Universit\'e de Toulouse, CNRS - Toulouse, France
\aff{2}
CNRS, UMR 5519, LEGI, F-38000 Grenoble, France
\aff{3} 
Univ. Grenoble Alpes, LEGI, F-38000 Grenoble, France
\aff{4}
Univ. Grenoble Alpes, Irstea, UR ETGR, 2 rue de la Papeterie-BP 76, F-38402 St-Martin-d'H\`eres, France
}

\maketitle

\begin{abstract}

Gravity-driven turbulent bedload transport has been extensively studied over the past century in regard to its importance for Earth surface processes such as natural riverbed morphological evolution. In the present contribution, the influence of the longitudinal channel inclination angle on gravity-driven turbulent bedload transport is studied in an idealised framework considering steady and uniform flow conditions. From an analytical analysis based on the two-phase continuous equations, it is shown that : (i) the classical slope correction of the critical Shields number is based on an erroneous formulation of the buoyancy force, (ii) the influence of the slope is not restricted to the critical Shields number but affects the whole transport formula and (iii) pressure-driven and gravity-driven turbulent bedload transport are not equivalent from the slope influence standpoint. 
Analysing further the granular flow driving mechanisms, the longitudinal slope is shown to not only influence the fluid bed shear stress and the resistance of the granular bed, but also to affect the fluid flow inside the granular bed - responsible for the transition from bedload transport to debris flow. The relative influence of these coupled mechanisms allows us to understand the evolution of the vertical structure of the granular flow and to predict the transport rate scaling law as a function of a rescaled Shields number. The theoretical analysis is validated with coupled fluid-discrete element simulations of idealised gravity-driven turbulent bedload transport, performed over a wide range of Shields number values, density ratios and channel inclination angles. In particular, all the data are shown to collapse onto a master curve when considering the sediment transport rate as a function of the proposed rescaled Shields number. 

\end{abstract}

\section{Introduction}

Turbulent bedload transport is of major importance for the prediction of riverbed evolution and coastal processes, which represent important issues for public safety, management of water resources and environmental sustainability. In this framework, the key parameter to predict is the dimensionless sediment transport rate \citep{Einstein1942}, $Q_s^*= Q_s/\sqrt{(\rho^p/\rho^f-1)g d^3}$, as a function of the dimensionless fluid bed shear stress denoted as the Shields number \citep{Shields1936}, $\theta^*= \tau_b/[(\rho^p - \rho^f)g d]$, where $Q_s$ is the volumetric sediment transport rate per unit width, $g$ is the acceleration of gravity, $d$ the particle's diameter, $\rho^p$ and $\rho^f$ are the particle and fluid densities and $\tau_b$ is the fluid bed shear stress. Due to the inherent complexity of granular media behaviour and turbulent fluid flows, turbulent bedload transport understanding remains limited despite a century of modern research on the subject \citep{Gilbert1914,Bagnold1956,Frey2011,Duran2012,Aussillous2013}. This is illustrated by the poor predictions provided by the classical formulas linking Einstein and Shields numbers - such as the \citet{MPM1948} formula - which lead to sediment transport rate up to two orders of magnitude different from what is observed in the field \citep{Recking2013}. Accordingly, the present paper focuses on the analysis of the slope influence in turbulent bedload transport, which might be one of the key aspect of the observed data dispersion. \\

Most applications of turbulent bedload transport involve the presence of a slope, for example in the case of a beach in coastal sediment transport, a river or a mountain stream. The slope inclination angle is expected to affect the sediment transport rate through a modification of the particle's mobility. This is classically accounted for by considering a force balance on a single grain at the top of the granular bed close to the onset of motion \citep{Fredsoe1992,Andreotti2013}. In the zero-slope case (see figure \ref{figSituationSlope}), the streamwise force balance at the onset of motion reduces to an equality between the streamwise force induced by the fluid flow and the resistive sliding friction force on the granular layer below. Considering only the main fluid forces to apply to the grain, i.e. drag and buoyancy \citep{Schmeeckle2007} the friction force can be expressed as a granular friction coefficient, $\mu_s$, multiplied by the buoyant weight representing the vertical force applied to the grain. Then, the force balance on a grain at the onset reads:
\begin{equation}
\frac{\pi}{8} \rho^f d^2 C_D u_*^2- \mu_s \left(\rho^p g ~ \frac{\pi}{6} d^3 - (f_b)_z\right)= 0,
\label{forceBalance}
\end{equation}
where $C_D$ is the drag coefficient, $\mathbf f_b$ is the buoyancy force and $u_*$ is the velocity scale at the granular bed assimilated to the fluid bed friction velocity. From this balance, by expressing the buoyancy force, the critical Shields number can be written as: 
\begin{equation}
\theta_c^0 =  \frac{\rho^f u_*^2}{ (\rho^p - \rho^f) gd } =  \frac{4 \mu_s}{3 C_D}. 
\label{modifNoSlope}
\end{equation}
In the presence of a longitudinal slope inclination angle, $\alpha$, (see figure \ref{figSituationSlope}) two additional positive terms appear in the force balance due to the projection of the particle weight and buoyancy force along the streamwise axis, while the friction force is reduced due to the projection of the particle weight along the vertical axis:
\begin{equation}
\frac{\pi}{8} \rho^f d^2 C_D u_*^2 -  \mu_s \left(\rho^p  g \cos \alpha  \frac{\pi}{6} d^3 - (f_b)_z \right) \  +  \rho^p g \sin \alpha \ \frac{\pi}{6} d^3 + (f_b)_x = 0.
\label{forceBalanceSlope}
\end{equation}
Taking the buoyancy force as  $\mathbf f_b = - \rho^f \frac{\pi}{6} d^3 \mathbf{g}$, leads to the following reduction of the critical Shields number with increasing slope:
\begin{figure}
  \centerline{  \includegraphics[width=0.48\textwidth]{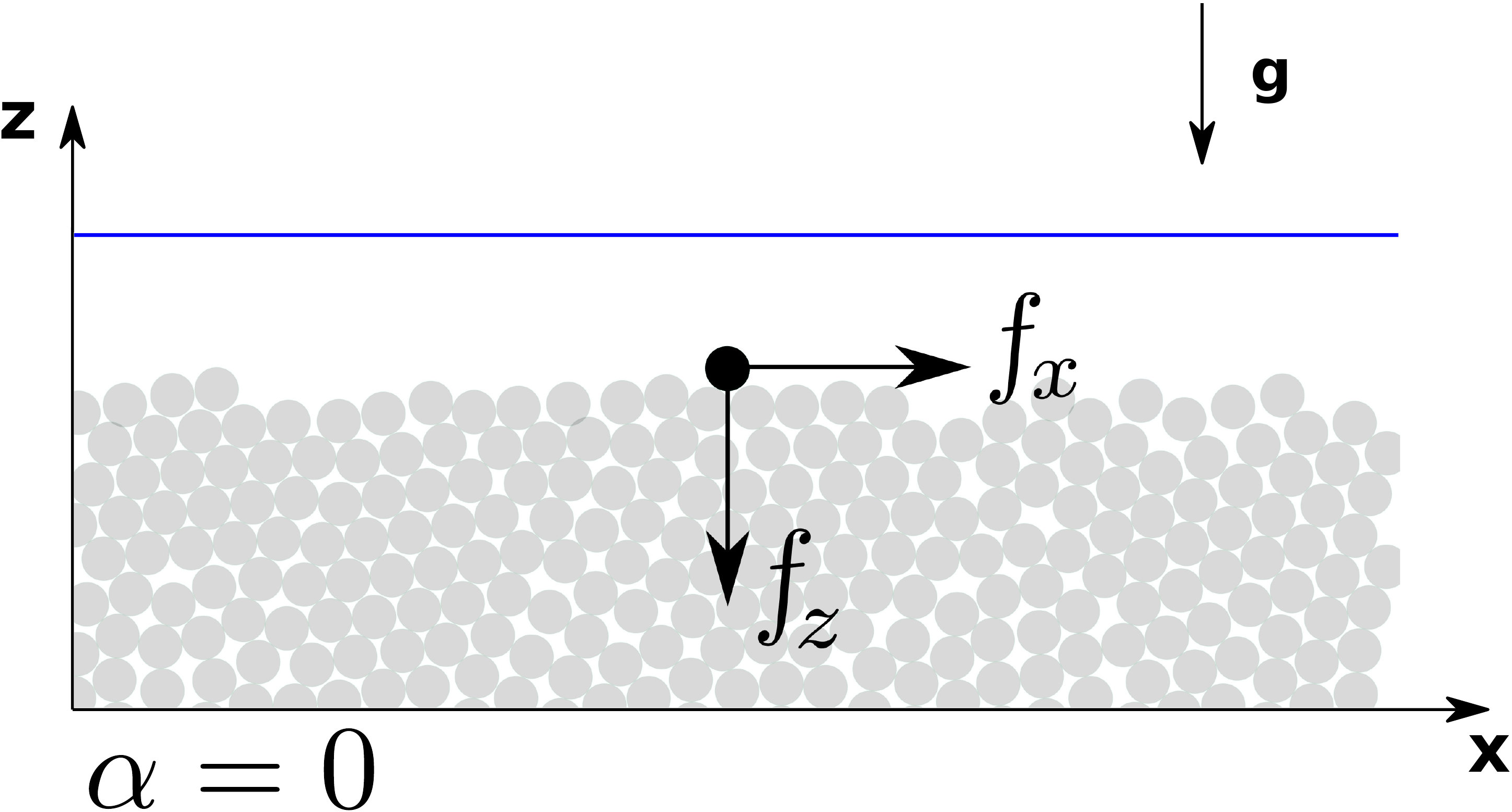}\includegraphics[width=0.48\textwidth]{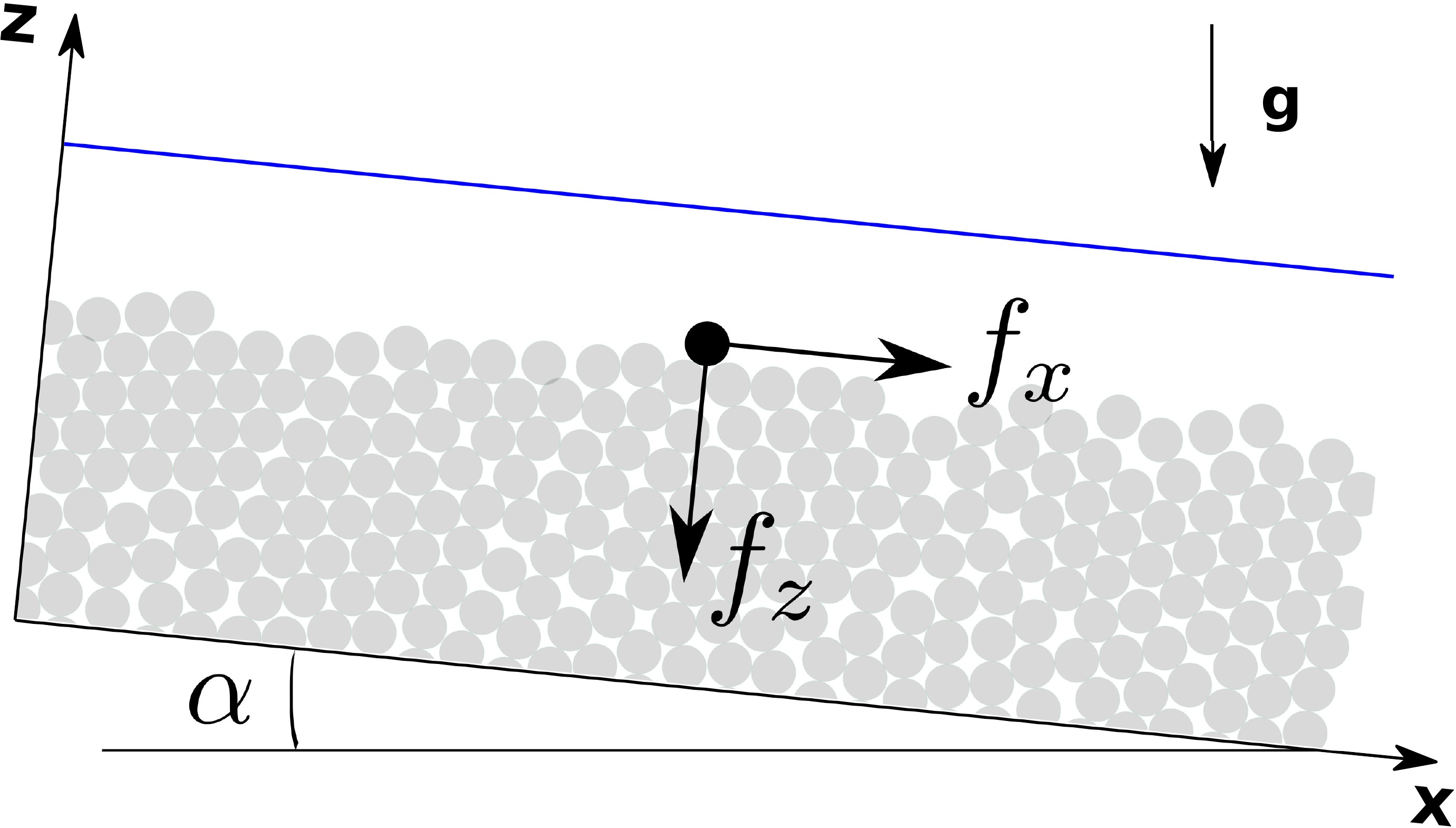}\\}
\caption{\label{figSituationSlope} Schematic slope influence on a particle at the top of the granular bed at rest. }
\end{figure}
\begin{equation}
\theta_c (\alpha)= \theta_c^0 \cos \alpha \left[ 1- \frac{\tan \alpha}{\mu_s} \right].\\
\label{modifTheta}
\end{equation}

This expression of the modification of the critical Shields number has been first formulated by \citet{Fernandez1976} considering the onset of motion to be due to rolling instead of sliding. The consecutive moment balance on the grain leads to the same expression of the modified critical Shields number, with $\mu_s$ the tangent of the so-called pocket angle formed by the local arrangement between the particle and its neighbours \citep{Wiberg1987}. Performing turbulent bedload transport experiments in an inclined rectangular pressure-driven closed conduit, with variation of particle diameter ($d \in [0.9,3.3]mm$), density ratio $\rho^p/\rho^f \in [1.34,4.5]$ and channel inclination angle ($\alpha \in [0,22]^{\circ}$), \citet{Fernandez1976} found a relatively good agreement between the theoretical prediction and experimental results providing a fit of the pocket angle, which appeared unexpectedly large. Following this pioneering work, \citet{Chiew1994} reproduced a similar approach, deriving equation (\ref{modifTheta}) from a force balance associating $\mu_s$ with the granular medium repose angle and comparing the prediction to experimental data in pipe flows. Varying the particle mean diameter $d \in [0.5,3]mm$, the repose angle $\Phi \in [33,38]^{\circ}$ and the channel inclination angle $\alpha \in [0,31]^{\circ}$, they showed that the critical Shields number follows the prediction of equation (\ref{modifTheta}) for downward slopes. This work has been further generalised by \citet{Seminara2002} and \citet{Dey2003} to the combined effect of transverse and longitudinal slopes, validating the analysis with experiments in pipe flows in the latter study. \\
This type of approach is widespread in turbulent sediment transport and can be found in classical textbooks of sediment transport \citep{Fredsoe1992} and granular media \citep{Andreotti2013}, as well as in the aeolian saltation community \citep{Iversen1994,Iversen1999}. It is also known in the morphodynamic community as the Ikeda-Coleman-Iwagaki model \citep{Wiberg1987} and has been applied in turbulent bedload transport to experiments and field transport rate predictions (see e.g. \citet{Li1999,Wilcock2003,Karmaker2016}). In addition, it has been extended to account for the lift force \citep{Wiberg1987,Chiew1994,Armanini2005b}, the fluid viscous sub-layer at low slope \citep{Wiberg1987} and the duration of a given applied fluid force in the context of turbulent fluid flows \citep{Diplas2008,Valyrakis2010}. \\ 

In the meantime, the longitudinal slope effect has been studied experimentally in gravity-driven turbulent bedload transport. Considering a large range of slopes,  \citet{Smart1983,Smart1984} and later \citet{Rickenmann1991,Rickenmann2001} have determined empirical relationships between the dimensionless sediment transport rate and a modified Shields number. The latter, classically evaluated from the water depth, $h$, was determined from the so-called mixture depth, $h_m = h+\delta_s$, including both the water depth, $h$ and the thickness of the granular layer in motion, $\delta_s$. As a consequence, the Shields number has a different meaning and this approach cannot be compared directly to the classical one, as the mobile layer thickness $\delta_s$ scales with the classical Shields number and the slope \citep{Sumer1996,Hsu2004,RevilBaudard2013}. These dependencies are expected to modify the classical scaling laws, underlining the importance of adopting the same definition of the Shields number when comparing experimental data and theoretical predictions. Although in a different framework, these studies show that the slope not only modifies the critical Shields number but also the Shields number definition. As a consequence, the empirical law relating the dimensionless sediment transport rate to the Shields number is also modified. A similar modification of the sediment transport formula has been proposed by \citet{Cheng2014}, replacing the gravity contribution by a slope-modified gravity in both the Shields and Einstein number formulations and considering the classical \citet{MPM1948} formula. The obtained formula fits the existing experimental data better than the classical corrections \citep{Cheng2014}, but lacks a solid theoretical justification. Lastly, it is interesting to note that \citet{Damgaard1997} also proposed an empirical correction of the transport formula based on experimental data in a closed conduit. \\

The literature review underlines two different trends that seem to be associated with pressure-driven and gravity-driven configurations in turbulent bedload transport. On one hand, gravity-driven experiments exhibit a modification of the transport formula as a function of the slope. On the other hand, the variation of the critical Shields number with the slope seems to be well predicted from a force/moment balance on a single particle in pressure-driven configurations. However, there is \textit{a priori} no theoretical justification why a variation of the slope should only affect the critical Shields number and the study of \citet{Damgaard1997} at moderate Shields number values suggests a behaviour similar to the gravity-driven configuration. This lack of characterisation together with the absence of clear theoretical bases in the literature suggests the need for further analysis. \\

In the present contribution, we attempt to give a better understanding of the longitudinal slope influence on turbulent bedload transport by adopting an idealised and theoretical point of view. Focusing on gravity-driven turbulent bedload transport under steady uniform flow conditions, we discuss the bases of the critical Shields number derivation and analyse the granular entrainment mechanisms in the framework of the two-phase continuous equations (section \ref{TheoreticalAnalysis}). This allows us to characterise the influence of the slope on the vertical flow structure and propose a re-scaling of the Shields number to account for the slope influence on the sediment transport rate. The proposed scaling is tested against fluid-discrete element method simulations (section \ref{simu}) and the theoretical results are discussed more generally in the light of the numerical results, considering in particular the vertical flow structure and the difference between gravity-driven and pressure-driven configurations (section \ref{discussion}).

\section{Theoretical analysis}
\label{TheoreticalAnalysis}

\subsection{Discussion on the classical critical Shields number derivation}
\label{Error}

The classical derivation of the critical Shields number reproduced in the introduction relies on the following expression of the buoyancy force applied to a particle:
\begin{equation} 
\mathbf f_b = - \rho^f \frac{\pi}{6} d^3 \mathbf{g}.
\label{buoyancyDefWrong}
\end{equation}
While this expression is classically used, it does not apply \textit{a priori} to all the different configurations explored in fluid-particle flow, as stressed by \citet{Christensen1995}.\\ The buoyancy force is defined as the force a fluid element would undergo if it was occupying the position of the particle \citep{Maxey1983}. It can be derived explicitly in the Stokes flow case and leads to the following formulation \citep{Maxey1983}:
\begin{equation}
\mathbf{f}_b = - \frac{\pi}{6} d^3 \ \nabla . \sigma_u^f,
\label{buoyancyDef}
\end{equation}
where $\sigma^f_u$ is the undisturbed fluid velocity stress tensor, i.e. the fluid stress tensor based on the fluid velocity field undisturbed by the presence of the particles. This formulation is used in \citet{Jackson2000} at the level of the continuous medium, including the Reynolds stresses contribution ($R^f$) inside the fluid stress tensor, i.e. replacing $\sigma_u^f$ by $\sigma_u^f + R^f$ inside equation (\ref{buoyancyDef}). However, the formulation of the buoyancy force should not \textit{a priori} include the contribution from the Reynolds stresses considering that they are associated with advection and related to the time averaging of the fluid velocity fluctuations. Indeed, considering a single particle in a turbulent Newtonian fluid flow, it experiences a time-averaged buoyancy force depending on the average fluid velocity field but independent of the fluctuations with time. This is the case because the expression of the buoyancy is linear with the second spatial derivative of the fluid velocity for a Newtonian fluid,
\begin{equation}
\mathbf{f}_b = - \frac{\pi}{6} d^3 \ \eta^f \ \nabla.[\nabla \mathbf{u^f} + (\nabla \mathbf{u^f})^T],
\end{equation}
with $\eta^f$ the dynamic fluid viscosity, so that the time averaging of the buoyancy force reduces to the contribution from the average fluid velocity field. From these considerations, the expression of the buoyancy force should follow the equation (\ref{buoyancyDef}) in all the different fluid flow regimes \citep{Lhuillier2011}.\\

As a consequence, the above-mentioned derivation is not valid in uniform gravity-driven turbulent bedload transport as the undisturbed fluid equation, $0 = -\nabla R^f  -\nabla \sigma_{u}^f + \rho^f \mathbf g$,  together with the regular expression of the buoyancy force (eq. \ref{buoyancyDef}), does not allow us to write the buoyancy force as equation (\ref{buoyancyDefWrong}).\\ Expressing the buoyancy directly from equation (\ref{buoyancyDef}), it follows that $(f_b)_z = \pi d^3/6 \rho^f g \cos \alpha$ and $(f_b)_x = - \pi d^3/6 ~ \partial_z \sigma^f_{xz}$, by using the hydrostatic characteristics of the fluid along $z$. Considering that the viscous shear stress contribution is negligible in turbulent bedload transport $\partial_z \sigma^f_{xz} \simeq 0$ \citep{RevilBaudard2013,Maurin2016}, the force balance performed on a particle lying on the granular bed at the onset of motion (eq. \ref{forceBalanceSlope}) reads:
 \begin{equation}
\frac{\pi}{8} \rho^f d^2 C_D u_*^2 -  \mu_s (\rho^p - \rho^f) g \cos \alpha ~\frac{\pi}{6} d^3  +  \rho^p g \sin \alpha ~ \frac{\pi}{6}d^3 = 0,
\label{eqOkForceBal}
\end{equation}
where only the particle weight projection along the streamwise axis is non-negligible. This leads to the following modification of the critical Shields number due to the slope effect:
\begin{equation}
\theta_c(\alpha) =   \theta_c^0\cos \alpha \left[1 - \frac{\tan \alpha}{\mu_s} \frac{1}{1-\rho^f/\rho^p} \right],
\end{equation}
where the density ratio $\rho^p/\rho^f$ modifies the classical slope effect characterised by equation (\ref{modifTheta}). This difference is of importance for underwater bedload transport, where the density ratio is typically of order one ($\rho^p \gtrsim \rho^f$). \\

Adopting the regular expression for the buoyancy force also affects the derivation of the slope influence on the critical Shields number in the pressure-driven configuration. In this case, the streamwise component of the buoyancy force is completed by the term due to the presence of a longitudinal pressure gradient:
\begin{equation}
\frac{\pi}{8} \rho^f d^2 C_D u_*^2 -  \mu_s (\rho^p - \rho^f) g \cos \alpha ~\frac{\pi}{6} d^3  +  \rho^p g \sin \alpha ~ \frac{\pi}{6}d^3 - \frac{\partial P^f}{\partial x} = 0,
\end{equation}
which does not enable us to recover the classical modification of the critical Shields number derived in the introduction (eq. \ref{modifTheta}). The latter equation remains only valid for laminar flows and in the case where the fluid mass column stays at a constant level perpendicular to the gravity, i.e. for underwater avalanches or coastal sediment transport. \\

\subsection{Vertical granular flow structure and entrainment mechanisms}
\label{GranularFlowStructure}

Instead of focusing on the critical Shields number modification in a discrete particle framework, we adopt a more general approach based on the continuous two-phase flow framework to analyse the slope influence from the onset of motion up to intense turbulent bedload transport. In order to better understand the vertical structure of the granular flow and the local mechanisms at play, let us express the shear-to-normal granular stress ratio as a function of the depth $\tau^p_{xz}(z)/P^p(z)$. The yield criterion  $\tau^p_{xz}/P^p>\mu_s$ is characteristic of the onset of granular flow, so that the positive contributions to the shear-to-normal stress ratio at a given height $z$ reflects the local entrainment mechanism of the granular medium.\\

The two-phase volume-averaged equations read for steady uniform flows \citep{Anderson1967,Jackson2000,Chauchat2017}:
\begin{equation}
0 = \frac{\partial \tau_{xz}^f}{\partial z} + \frac{\partial R_{xz}^f}{\partial z} + \rho^f (1-\phi) g \sin \alpha - n \left<{f_f^p}_x\right>^p,
\label{fluidXZ}
\end{equation}
\begin{equation}
0 = \frac{\partial \tau_{xz}^p}{\partial z} + \rho^p \phi  g \sin \alpha + n \left<{f_f^p}_x\right>^p,
\label{partXZ}
\end{equation}
\begin{equation}
0 = - \frac{\partial P^f}{\partial z} + \rho^f (1-\phi)  g \cos \alpha - n \left<{f_f^p}_z\right>^p,
\label{fluidZZ}
\end{equation}
\begin{equation}
0 =  - \frac{\partial P^p}{\partial z} + \rho^p \phi  g \cos \alpha  + n \left<{f_f^p}_z\right>^p,
\label{partZZ}
\end{equation}
Where $\sigma_{ij}^f = -P^f \delta_{ij} + \tau_{ij}^f$ is the effective fluid stress tensor with $\delta_{ij}$ the Kronecker's delta, $\sigma_{ij}^p = -P^p \delta_{ij} + \tau_{ij}^p$ is the granular stress tensor, $R^f_{ij}$ is the Reynolds stress tensor, $\phi$ and $\epsilon$ are the solid and fluid volume fraction respectively and $ n \left<{f_f^p}_k\right>^p$ is the $k$ component of the fluid-solid momentum transfer, with $n$ the particle density and $\left<{f_f^p}_k\right>^p$ the volume-averaged fluid-particle interaction force. These equations represents the volume-averaged momentum balance over the streamwise and wall-normal directions of the fluid and solid phases and the brackets with superscript $p$ denotes the classical volume averaging over the particle phase defined in \citet{Anderson1967}.\\
Integrating these equations between a position $z<h_p$ in the moving granular layer and $h_p$, the top of it, allows us to simplify the equations and to express the shear-to-normal stress ratio as a function of the vertical position $z$. Focusing on the granular equations (\ref{partXZ}) and (\ref{partZZ}), the granular stress tensor is zero at $h_p$ and the integrated  streamwise component of the granular momentum balance leads to: 
\begin{equation}
\tau_{xz}^p(z) = \rho^p  g \sin \alpha \displaystyle \int_{z}^{h_p}{\phi(\zeta) d\zeta} + \displaystyle \int_{z}^{h_p}{n \left<{f_f^p}_x\right>^p(\zeta) d \zeta}.
\label{eq1}
\end{equation}
Assuming that the fluid wall-normal pressure is hydrostatic, the fluid-particle interaction term $n \left<{f_f^p}_z\right>^p$ can be expressed from equation \ref{fluidZZ}. Replacing it in equation (\ref{partZZ}), we obtain:
\begin{equation}
P^p(z) = g \cos \alpha(\rho^p-\rho^f)\displaystyle \int_{z}^{h_p}{\phi(\zeta) d\zeta}.
\label{eq2}
\end{equation}
Lastly, the fluid streamwise momentum balance (eq. \ref{fluidXZ}) can be  integrated between $z$ and $h_p$ in order to express the last term on the right-hand side of equation (\ref{eq1}):
\begin{equation}
0 = \tau_{xz}^f(h_p)-\tau_{xz}^f(z) + R_{xz}^f(h_p) -R_{xz}^f(z) + \rho^f g \sin \alpha \displaystyle \int_{z}^{h_p}{[1-\phi(\zeta)]d \zeta} -  \displaystyle \int_{z}^{h_p}{n \left<{f_f^p}_x\right>^p(\zeta) d \zeta}.
\label{eq2}
\end{equation}
It has been shown previously in turbulent bedload transport and sheet-flow simulations that the effective viscous shear stress tensor is negligible throughout the depth with respect to the Reynolds stresses and to the slope contribution \citep{RevilBaudard2013,Maurin2016}. Furthermore, the Reynolds stresses are fully damped in the granular bed and we assume that their contribution is negligible in the moving granular layer $R_{xz}^f(z) \sim 0$. Also, the fluid bed shear stress used to define the Shields number is taken here as the maximum of the turbulent shear stress \citep{RevilBaudard2013,Maurin2016}, assuming that the latter is located at the top of the granular layer, $\tau_b = R_{xz}^f(h_p)$. Therefore, equation (\ref{eq2}) reads: 
\begin{equation}
 \displaystyle \int_{z}^{h_p}{n \left<{f_f^p}_x\right>^p(\zeta) d \zeta} = \tau_b + \rho^f g \sin \alpha \displaystyle \int_{z}^{h_p}{[1-\phi(\zeta)]d \zeta}.
\label{eq3}
\end{equation}
Combining equations (\ref{eq1}), (\ref{eq2}) and (\ref{eq3}) and defining the average solid volume fraction between $z$ and $h_p$, $\bar{\phi}_z$ as:
\begin{equation}
\bar{\phi}_z (h_p - z) = \displaystyle \int_{z}^{h_p}{\phi(\zeta) d \zeta},
\end{equation}
the shear-to-normal stress ratio $\tau_{xz}^p(z)/\tau_{zz}^p(z)$ can be expressed as:
\begin{equation}
\displaystyle \frac{\tau_{xz}^p(z)}{\tau_{zz}^p(z)}= \frac{\rho^p}{\rho^p-\rho^f} \tan \alpha + \frac{ \tau_b}{g \cos \alpha(\rho^p-\rho^f)\bar{\phi}_z (h_p - z)} +\frac{ \rho^f}{\rho^p-\rho^f} \frac{1-\bar{\phi}_z}{\bar{\phi_z}} \tan \alpha,
\label{stressRatioFinal}
\end{equation}

\vspace{0.5cm}

Three contributions to the granular phase shear-to-normal stress ratio can be identified. The first term on the right-hand side of equation (\ref{stressRatioFinal}) is constant within the depth and represents the slope effect on the granular phase. It is analogous to the classical configuration of a dry granular medium on an inclined plane, where the shear-to-normal stress ratio is constant within the depth and equal to the tangent of the inclination angle $\alpha$ \citep{Andreotti2013}. In bedload transport, the presence of a buoyancy force along the vertical axis leads to a modification of this term by the density ratio. The two last terms on the right-hand side correspond to the fluid contribution on the granular phase, which has been split into two contributions coming from the fluid flow above the bed and the fluid flow inside the granular layer respectively. The Shields number can be identified in the former and the term reduces to: 
\begin{equation}
\frac{ \tau_b}{g \cos \alpha(\rho^p-\rho^f)\bar{\phi}_z (h_p - z)} = \frac{\theta^*}{\bar{\phi}_z}\frac{d}{h_p - z}.
\end{equation}
This represents the contribution from the fluid bed shear stress to the entrainment of particles. Its surface nature is characterised by a decrease of this contribution inside the granular bed, i.e. as $z$ decreases. It is accounted for in bedload transport studies, by assuming that the Shields number is representative of the total imposed fluid shear stress. However, the contribution from the fluid flow inside the granular layer, corresponding to the third term on the right-hand side of equation (\ref{stressRatioFinal}), is usually neglected while it is seen to be of importance when increasing the slope angle. This contribution depends on the slope and characterises the effect of the slope on the fluid flow inside the granular layer. The latter affects the granular phase through the fluid-particle interaction and in consequence depends also on the specific density. In addition, the different dependence on the vertical position $z$ of the two fluid terms implies that a modification of the slope at given Shields number and specific density would induce a change in the vertical repartition of the shear-to-normal stress ratio and therefore on the vertical structure of the granular flow. This could explain why the velocity and solid volume fraction profiles are not self-similar in gravity-driven bedload transport \citep{Armanini2005,Larcher2007,Capart2011,Frey2014,MaurinPhD,Maurin2015}.\\

In order to gain more insight into the physical  meaning of the fluid flow contribution inside the granular bed, let us consider the simple case of a gravity-driven fluid flow through a quasi-static porous granular bed. The Shields number contribution is negligible in this configuration and the granular bed is at rest so that the average solid volume fraction is equal to the maximum packing fraction $\bar{\phi}_z = \phi^{max}$. This leads to a shear-to-normal stress ratio independent of the vertical position $z$:
\begin{multline}
\displaystyle \frac{\tau_{xz}^p(z)}{P^p(z)}= \frac{\rho^p}{\rho^p-\rho^f} \tan \alpha +\frac{\rho^f}{\rho^p-\rho^f} \frac{1-\phi^{max}}{\phi^{max}} \tan \alpha \\
 =\frac{\tan\alpha}{\rho^p/\rho^f - 1}\left[\frac{\rho^p}{\rho^f} +  \frac{1-\phi^{max}}{\phi^{max}}\right].
\label{stressRatioBed}
\end{multline}
As a consequence, there exists a critical angle above which the entire granular layer will be entrained independently from its thickness. This avalanche angle, $\alpha_0$, corresponds to the configuration for which the shear-to-normal stress ratio exceeds the static friction coefficient of the granular medium $\mu_s$:
\begin{equation}
\displaystyle \tan \alpha_0  = \frac{\mu_s}{1+ \left[(\rho^p/\rho^f-1)\phi^{max}\right]^{-1}}.
\label{critDebrisAngle}
\end{equation}
This corresponds to the critical angle for the onset of debris flow as derived by \citet{Takahashi1978} and is well known in the debris flow community \citep{Takahashi2007}. The derivation presented here makes the link between debris flows and gravity-driven turbulent bedload transport and evidences the importance of the slope influence through the fluid flow inside the granular layer when considering steep slope configurations. In addition, it shows that the expression of the critical angle of debris flow $\alpha_0$ is characteristic of the slope influence on the granular phase through both its direct impact and the fluid-induced contribution in gravity-driven bedload transport, two mechanisms that depend on the specific density $\rho^p/\rho^f -1$.

\subsection{Sediment transport rate scaling}
\label{subSecScaling}

Taking advantage of the developed analysis, we look for a scaling law of the sediment transport rate accounting for the slope effect. Following \citet{Bagnold1956}, the sediment transport rate per unit width $Q_s$ can be expressed as a function of the solid volume fraction $\phi$ and the average particle velocity  $\left<v_x\right>^p$ profiles: 
\begin{equation}
Q_s = \int_0^{\infty} \phi(z) \left<v_x\right>^p(z) dz = \int_{h_c}^{h_p} \phi(z) \left<v_x\right>^p(z) dz =  \overline{\phi \left<v_x\right>^p} \ \delta_s. 
\label{sedimentRate}
\end{equation}
where $h_c$ is the lower bound of the mobile granular layer, $\delta_s = h_p - h_c$ is the mobile granular layer thickness and $\bar{\bullet}$ denotes the average over $z$ along the mobile granular layer thickness. Considering $z = h_c$ in equation  (\ref{stressRatioFinal}), the mobile layer thickness can be expressed as: 
\begin{equation}
\displaystyle \frac{\delta_s}{d} = \frac{\theta^*}{\bar{\phi}\left[\mu_s - \left(1+ [\bar{\phi}(\rho^p/\rho^f -1)]^{-1} \right)\tan \alpha \right]} \simeq \frac{\theta^*}{\phi^{max} \mu_s \left[1- \tan \alpha /\tan \alpha_0 \right]}.
\label{mobilizedExtent}
\end{equation}
where equation (\ref{critDebrisAngle}) has been used and it has been assumed that $\bar{\phi}\simeq \phi^{max}$ as the average solid volume fraction $\bar{\phi}$ is only weakly varying in the problem.\\

The term $\overline{\phi \left<v_x\right>^p}$ has the dimension of a velocity, i.e. the square root of the product of an acceleration and a length scale. In the problem, the only acceleration scale is the buoyant reduced gravity $\tilde{g} = (\rho^p/\rho^f -1)g$ \citep{Duran2012} and we have seen that the relevant length scale is the mobile granular layer thickness. Thus, the volume flux should scale as: 
\begin{equation}
\overline{\phi \left<v_x\right>^p} \sim \sqrt{(\rho^p/\rho^f -1)g \delta_s}. 
\label{velocityScale}
\end{equation}
\vspace{0.2cm}
Compiling equations (\ref{sedimentRate}), (\ref{mobilizedExtent}) and  (\ref{velocityScale}), the scaling law of the dimensionless sediment transport rate $Q_s^*$ can be written as:
\begin{equation}
Q_s^* = \frac{Q_s}{\sqrt{(\rho^p/\rho^f -1)g d^3}} \sim \frac{\sqrt{(\rho^p/\rho^f -1)g \delta_s^3}}{\sqrt{(\rho^p/\rho^f -1)g d^3}} \sim \left(\frac{\theta^*}{\phi^{max} \mu_s \left[1- \tan \alpha/\tan \alpha_0 \right]}\right)^{3/2}.
\end{equation}

\vspace{0.5cm}

This approach not only allows us to recover the scaling law of the dimensionless sediment transport rate with the Shields number and the power $3/2$ in the high Shields number limit, but also allows us to express the influence of the slope through an effective granular friction coefficient, $\mu_{eff} = \mu_s \left[1- \tan \alpha/\tan \alpha_0 \right]$. The proposed scaling law is consistent with the original definition of the Shields number as a ratio between the driving and resistive forces applied to the granular phase. Indeed, it represents the ratio between the tangential fluid bed shear stress $\tau_b$ and the tangential shear stress associated with the friction between granular layers $\tau = \mu_{eff} P^p = \phi \mu_{eff} (\rho^p - \rho^f) g d$. The dependence of the effective friction coefficient on the distance to the critical debris flow angle, $1-\tan \alpha/\tan \alpha_0$, is consistent with the analysis of section \ref{GranularFlowStructure} and allows us to take into account both the direct and indirect slope effects on the granular phase in gravity-driven bedload transport. It suggests that the key parameter is not the slope in itself but the ratio between the slope and the critical angle of debris flow. This is of particular importance as the latter is not constant and depends on the maximum solid volume fraction and on the specific density. 

\section{Numerical Analysis}
\label{simu}

In order to verify the developed analysis, numerical simulations of turbulent bedload transport are performed using an existing coupled fluid-discrete element model \citep{Maurin2015}. 

\begin{figure}
  \centerline{  \includegraphics[width=\textwidth]{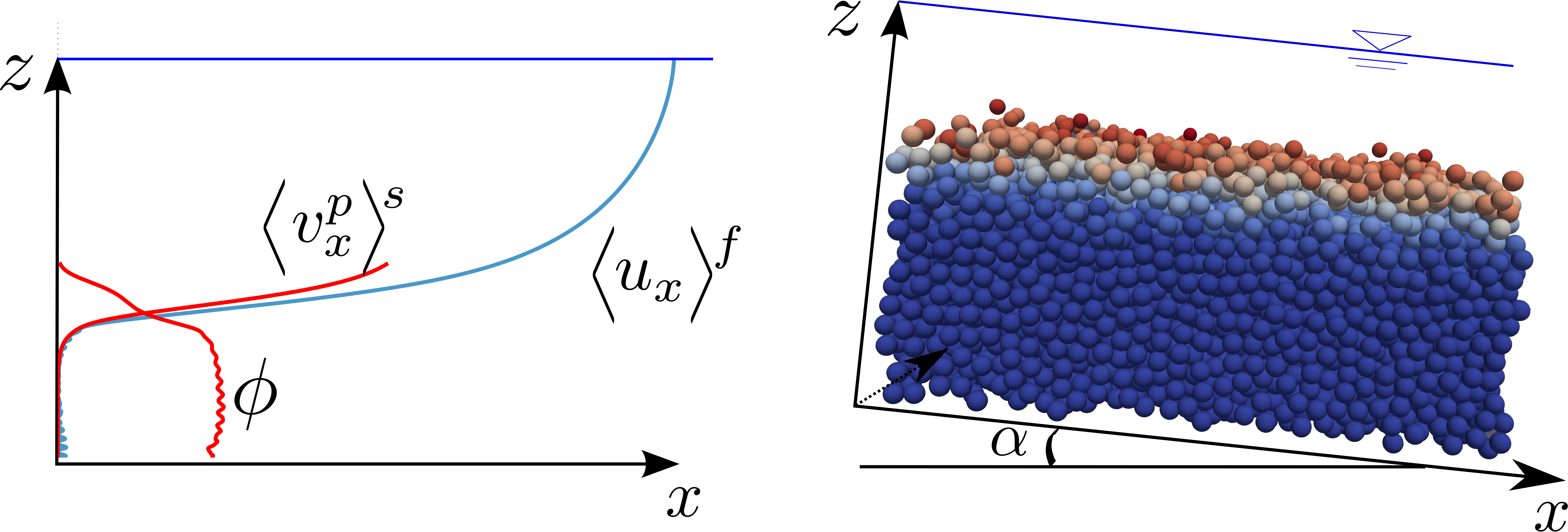} }
\caption{\label{figSituation} Scheme of the numerical setup and its equivalent average unidirectional picture with typical fluid velocity $\left<\mathbf{u}^f\right> = \left<u_x^f\right>(z) \ \mathbf{e_x}$, solid volume fraction $\phi$ and solid velocity $\left<\mathbf{v}^p\right>^s = \left<v_x^p\right>^s(z) \ \mathbf{e_x}$ depth profiles. The inclined 3-D bi-periodic granular description is coupled with a unidirectional fluid momentum balance using imposed fixed random bottom boundary condition and water free surface position. The particle color is representative of the velocity intensity. }
\end{figure}

\subsection{Numerical model}
The model is based on a three-dimensional (3-D) discrete element method (DEM) coupled with a 1-D volume-averaged fluid momentum balance. The fine resolution of the granular phase and the two-way coupling ensures an accurate description of the vertical depth profiles and the momentum conservation of the system on average. The model has been described in detail and validated experimentally in \citet{Maurin2015}, so that only the main characteristics are recalled here. The typical configuration considered here is shown in figure \ref{figSituation}.\\

The bi-periodic 3-D DEM is based on the explicit resolution of the particle phase, solving Newton's equations of motion for each particle $p$ at position $x_p$:
\begin{equation}
m \frac{d^2 \mathbf{x}^p}{d t^2} =   \mathbf{f}_{ext}^p + \sum_{k \in {\cal N}}\mathbf{f}_c^{pk} = \mathbf{f}_g^p + \mathbf{f}_f^p + \sum_{k \in {\cal N}}\mathbf{f}_c^{pk}
\label{DEM}
\end{equation}
where the sum of the contact forces $\mathbf{f}_{c}^{pk}$ is made over the ensemble of nearest neighbours ${\cal N}$, $\mathbf{f}_g^p$ is the gravity force, $\mathbf{f}_f^p$ is the force applied by the fluid on particle $p$.  Similarly, the rotation of the particles are solved from Newton's equations of motion. The contact force between particles is determined from the particles overlap using the classical spring-dashpot contact law \citep{Schwager2007} which defines a unique normal restitution coefficient $e_n$ and considers tangential friction characterised by a friction coefficient $\mu_p$. The latter two are taken as respectively $e_n = 0.5$ and $\mu_p= 0.4$, as determined from experimental comparisons \citep{Maurin2015}. The interaction with the fluid phase ($\mathbf{f}_f^p$) is mainly restricted to buoyancy, drag and lift forces in turbulent bedload transport \citep{Nino1994,Nino1998}. While it is clear that the lift force plays a non-negligible role in turbulent bedload transport \citep{Ji2013}, its expression has only been derived in the limits of Stokes flow \citep{Saffman1965} or an inviscid fluid \citep{Schmeeckle2007} and does not apply to the fluid flow regimes associated with turbulent bedload transport \citep{Schmeeckle2007}. As a consequence, it has been decided to avoid including a controversial expression of the lift force, which appeared to be unnecessary to reproduce accurately turbulent bedload transport experiments \citep{Maurin2015}. Therefore, the fluid-particle interactions are restricted in the present work to the three-dimensional buoyancy ($\mathbf{f}_{b}^p$) and drag forces ($\mathbf{f}_{D}^p$) \citep{Jackson2000,Maurin2015}:
\begin{equation} 
\mathbf{f}_{b}^p  =  -\frac{\pi d^3}{6}\mathbf{\nabla}P^f,
\label{generalisedBuoyancy}
\end{equation}
\begin{equation}
\displaystyle \mathbf{f}_{D}^p = \frac{1}{2}\rho^f \frac{\pi d^2}{4} ~ C_D ~ \left|\left| \left<\mathbf{u}\right>^f_{\mathbf{x^p}} - \mathbf{v^p} \right|\right|\left(\left<\mathbf{u}\right>^f_{\mathbf{x^p}} - \mathbf{v^p}\right),
\label{drag}
\end{equation}
where the average fluid velocity and the fluid pressure are taken at the centre of particle $p$ and $\mathbf{v^p}$ represents the velocity of particle $p$. The drag coefficient $C_D$ depends on the particle Reynolds number $\Rey_p = \left|\left| \left<\mathbf{u}\right>^f_{\mathbf{x^p}} - \mathbf{v^p} \right|\right| d/\nu^f$ and takes into account hindrance effects \citep{Dallavalle1948,Richardson1954}: $C_D = \left(0.4+ 24.4/Re_p\right) (1-\phi)^{-3.1}$. Knowing the position and velocity of the particles, the fluid pressure and velocity fields, Newton's equations of motion are solved for the ensemble of particles using the open-source DEM code YADE \citep{YADEDEM2015}. \\

The coupled fluid description solves the volume-averaged momentum balance of the fluid phase, corresponding to equations (\ref{fluidXZ}) and (\ref{fluidZZ}):
\begin{equation}
0 = \frac{\partial \tau_{xz}^f}{\partial z} + \frac{\partial R_{xz}^f}{\partial z} + \rho^f (1-\phi) g \sin \alpha - n \left<{f_f^p}_x\right>^p,
\label{fluidXZB}
\end{equation}
\begin{equation}
0 = - \frac{\partial P^f}{\partial z} + \rho^f (1-\phi)  g \cos \alpha - n \left<{f_f^p}_z\right>^p,
\label{fluidZZB}
\end{equation}
Due to the steady uniform character of the problem, the fluid velocity field reduces to its streamwise component and depends only on the vertical position: $\left<\mathbf{u}\right>^f = \left<u_x\right>^f(z) \ \mathbf{e_x}$, as sketched in figure \ref{figSituation}. Therefore, the buoyancy force (eq. \ref{generalisedBuoyancy}) is restricted to its wall-normal component and equation (\ref{fluidZZB}) leads to a hydrostatic fluid pressure distribution, the wall-normal average drag force being negligible. The solution of the streamwise fluid phase momentum balance (eq. \ref{fluidXZB}) requires closures for the viscous and Reynolds stress tensors, as well as the determination of the solid volume fraction $\phi$ and the momentum transfer associated with the granular phase interactions. Considering a Newtonian fluid, the effective viscous shear stress is expressed as: 
\begin{equation}
\tau_{xz}^f = \rho^f (1-\phi) \nu^f \frac{d \left<u_x\right>^f}{dz},
\label{viscousStressTensorClosure}
\end{equation}
with $\nu^f$ the clear fluid kinematic viscosity and $\left<u_x\right>^f$ the volume-averaged streamwise fluid phase velocity. The Reynolds shear stress is based on the eddy viscosity concept ($\nu^t$) using a mixing length formulation: 
\begin{equation}
 R_{xz}^f =  \rho^f ~ \nu^t \frac{d \left<u_x\right>^f}{dz} \\ \text{ with }\\ \nu^t = (1-\phi)  \ l_m^2 \left|\frac{d \left< u_x \right>^f}{dz}\right|,
\label{ReynoldsStressTensorClosure}
\end{equation}
where the mixing length is taken similarly to \citet{Li1995} as:
\begin{equation}
l_m(z) = \kappa \int_0^z{\frac{\phi^{max} - \phi(\zeta)}{\phi^{max}} ~d\zeta},
\label{mixingLength}
\end{equation}
with $\kappa=0.41$, the von K\'arm\'an constant. The formulation adopted allows us to recover the law of the wall \citep{Prandtl1926} in a clear fluid, while completely damping the turbulence inside the granular bed at maximum packing fraction ($\phi^{max}$).\\
The solid volume fraction ($\phi$) and the fluid-particle interaction term ($n \left<{f_f^p}\right>^p$) are determined from spatial averaging of the discrete solid phase. Considering cubic boxes of finite wall-normal length scale $l_z$, these two terms are averaged over the whole length and width of the granular bi-periodic cell. In order to solve the important wall-normal gradients present in turbulent bedload transport, a small wall-normal weighting function length scale has been adopted (typically $l_z = d/30$) and this choice has been confirmed by the experimental validation \citep{Maurin2015}.\\

The 3-D DEM and the fluid model are solved as transient problems applying a fixed bottom boundary condition for both the fluid ($\left<\mathbf{u}\right>^f(z = 0) = 0$) and the particle phase (fixed random particles) and imposing the position of the water free surface ($d \left<u_x\right>^f/dz (z = h) = 0$). In order to achieve a stable integration, the DEM time step is bounded by the propagation time of the fastest wave over a particle diameter \citep{MaurinPhD,Maurin2015}. The fluid resolution time step corresponds to a typical characteristic evolution time scale of the granular medium and is taken much larger than the DEM one \citep{Maurin2015}: $\Delta t_f = 10^{-2}s$ with respect to $\Delta t_p \sim O(10^{-4}-10^{-5}) \ s$.\\

The model has been compared with experiments and has shown its capability to describe accurately both the sediment transport rate and the granular depth structure in turbulent bedload transport \citep{Maurin2015}.

\subsection{Results}

In order to study the bulk flow behaviour and to investigate the parameter space, bi-periodic three-dimensional numerical simulations are performed by varying the Shields number $\theta^*$ between the onset of motion $\theta^*_c$ and intense bedload transport ($\theta^* \sim 0.6$), the relative slope angle $\tan \alpha /\tan \alpha_0$ between $0.1$ and $0.75$ and the specific density $\rho^p/\rho^f -1$ between $0.75$ (lightweight plastic), $1.5$ (glass/natural sediment) and $3$ (metal). Detailed parameters of the runs are shown in table \ref{tableParam}. In the simulations, the water free surface position and the channel slope angle are imposed before letting the system evolves under the effect of gravity. After reaching the steady state, the data are averaged over space every typical granular evolution time scale, $0.1s$, over 1000 measurements \citep{Maurin2015}. Note that the eddy turnover time varies between $0.01s$ and $0.1s$ in the configurations considered. Consistently with the theoretical derivation of section \ref{GranularFlowStructure}, the Shields number is evaluated by taking the fluid bed shear stress as the maximum of the time-averaged turbulent Reynolds shear stresses. The sediment transport rate is evaluated from equation (\ref{sedimentRate}), neglecting therefore the local instantaneous lateral and vertical contributions. The latter have been checked to be negligible with respect to the streamwise contribution. \\

\begin{figure}
  \centerline{  \includegraphics[width=0.5\textwidth]{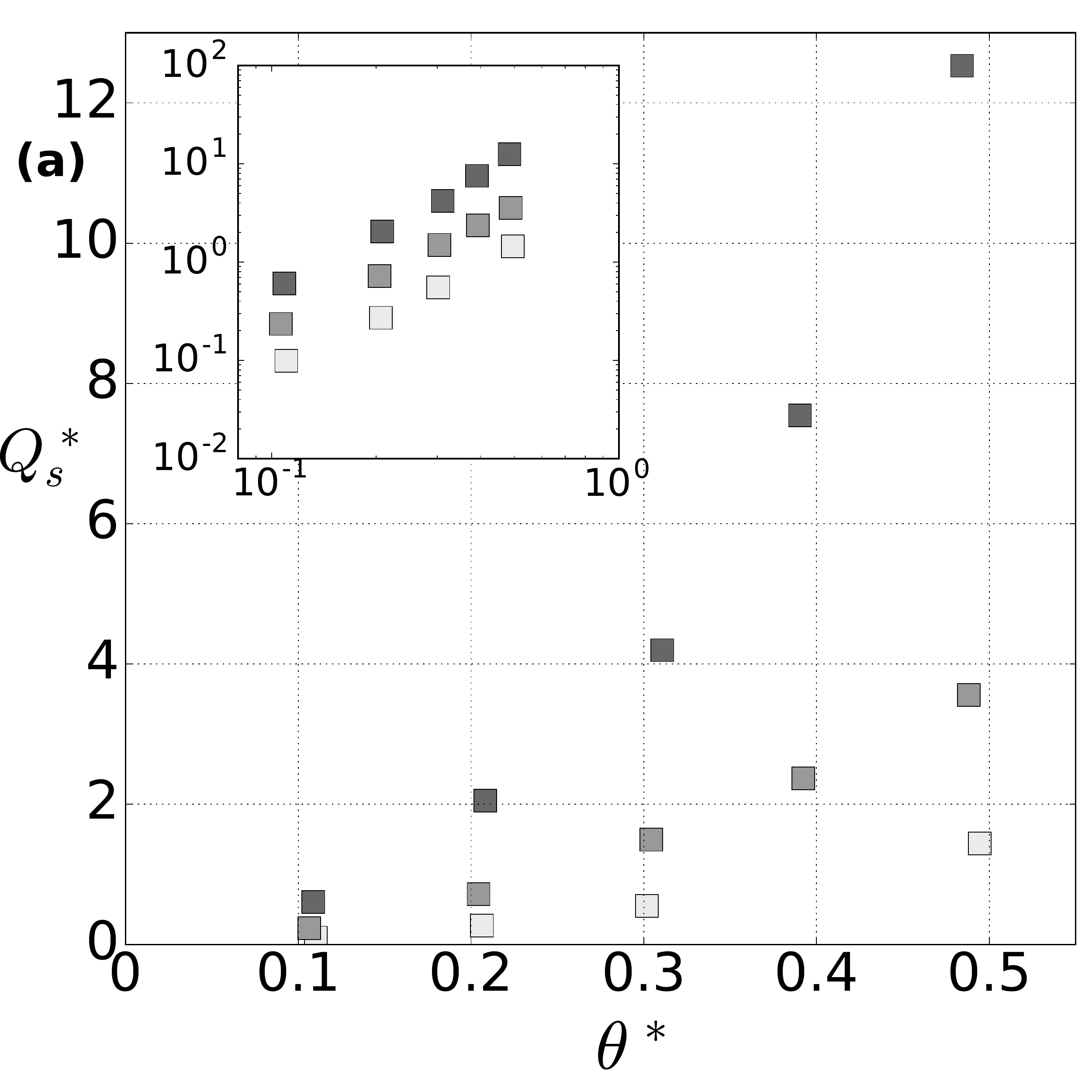}\includegraphics[width=0.5\textwidth]{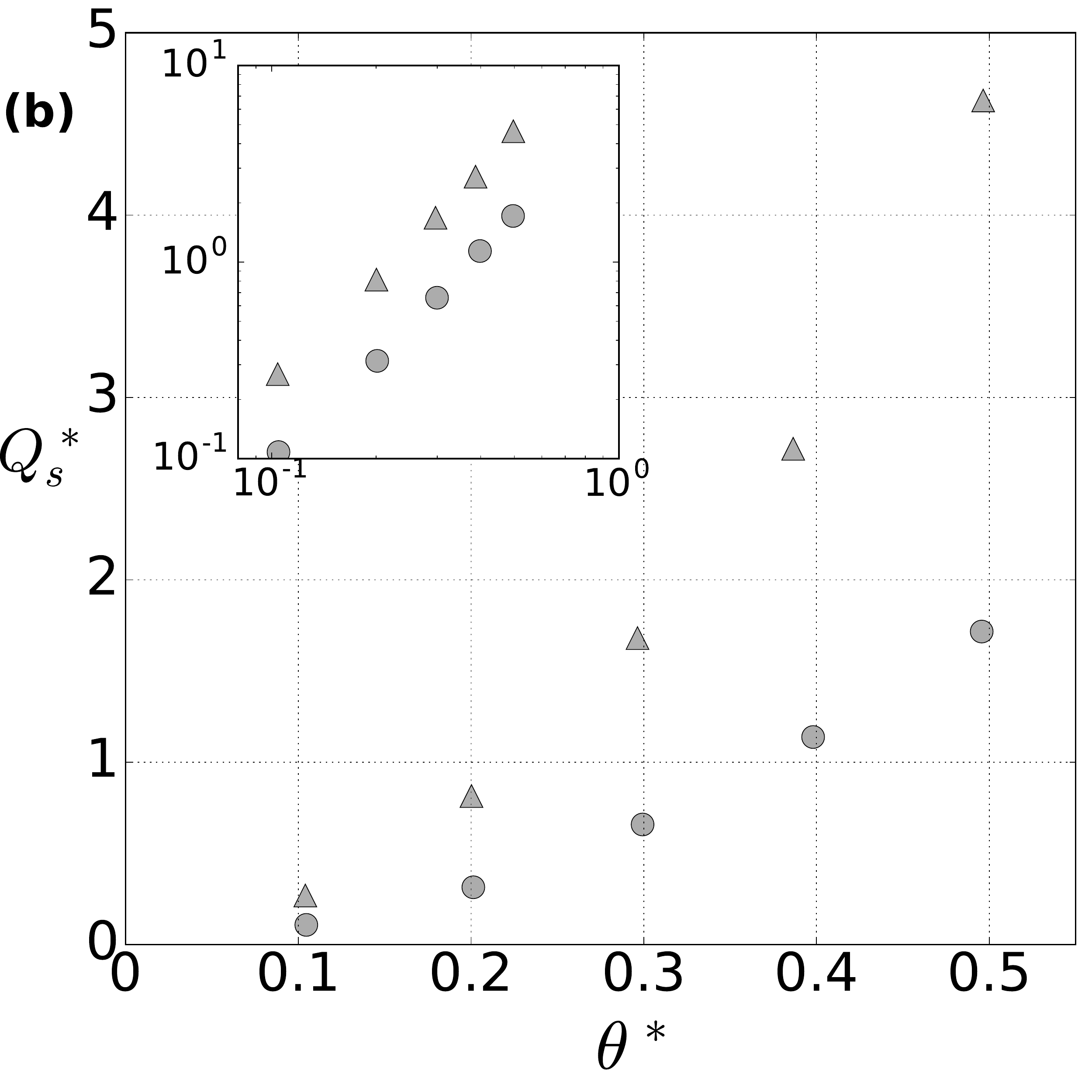} }
\caption{\label{slopeAndDensity} Dimensionless sediment transport rate $Q_s^*$ as a function of the Shields number $\theta^*$ for (a) variation of slope at a specific density of $\rho^p/\rho^f -1 = 1.5$ and (b) variation of specific density at a given slope. The triangles, the squares and the circles denote simulations with specific density $\rho^p/\rho^f -1$ of $0.75$, $1.5$ and $3$ respectively. The darkness of the points is characteristic of the value of the slope angle $\alpha$ varied between 0.02, 0.1 and 0.14 (1, 6 and 9 degrees).}
\end{figure}

Figure \ref{slopeAndDensity}(a,b) shows the dimensionless sediment transport rate as a function of the Shields number for variation of the channel inclination angle and the specific density respectively. While being far from the threshold of motion, an increase of up to an order of magnitude is observed in the dimensionless sediment transport rate when the slope increases at constant Shields number. This important influence at high Shields number evidences the impact of the bed slope on the sediment transport rate. In addition, the variation of the specific density at constant slope and Shields number is also seen to affect the dimensionless sediment transport rate (see fig. \ref{slopeAndDensity}b), as expected from the theoretical analysis presented in section \ref{TheoreticalAnalysis}.\\

\begin{figure}
  \centering  \includegraphics[width=0.5\textwidth]{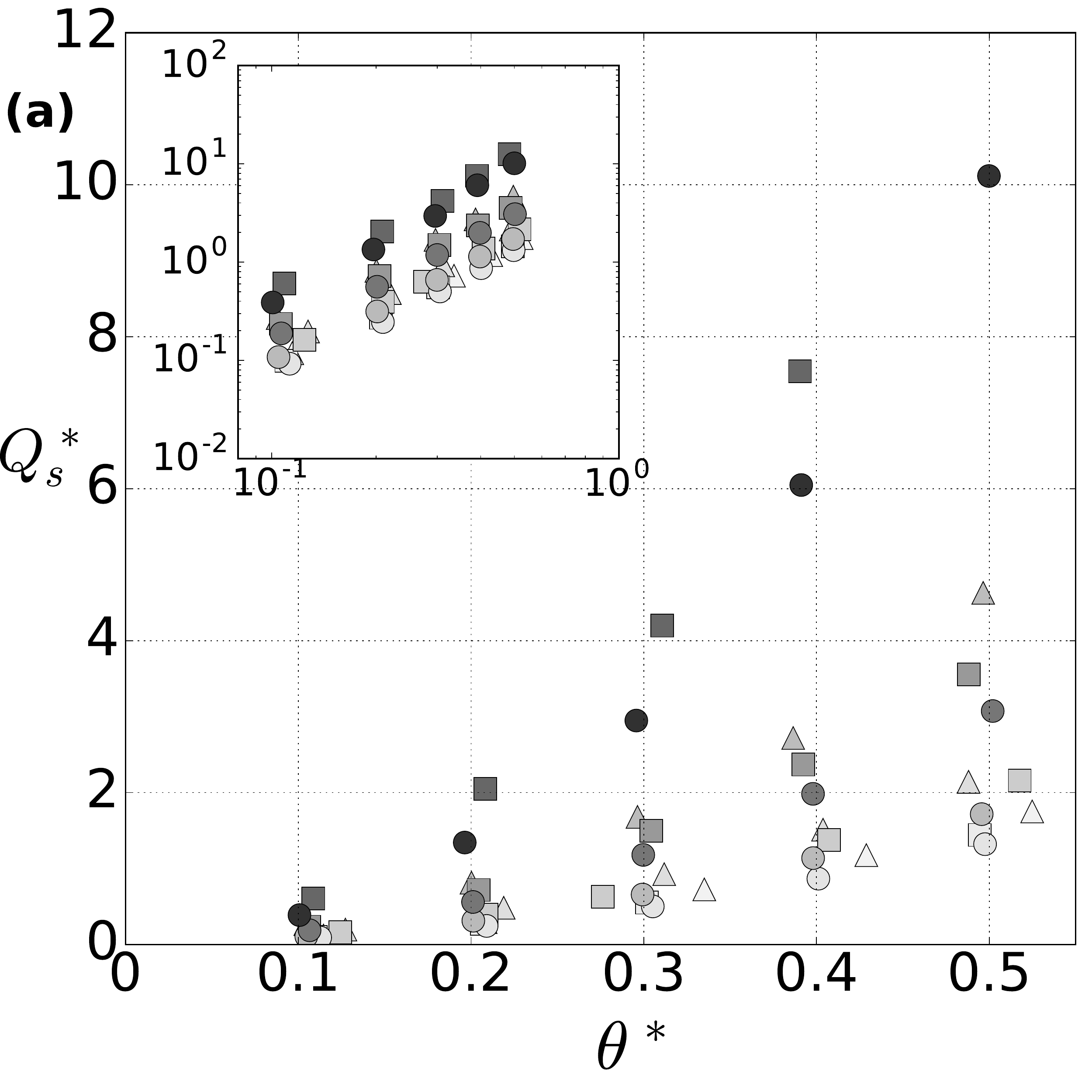}\includegraphics[width=0.5\textwidth]{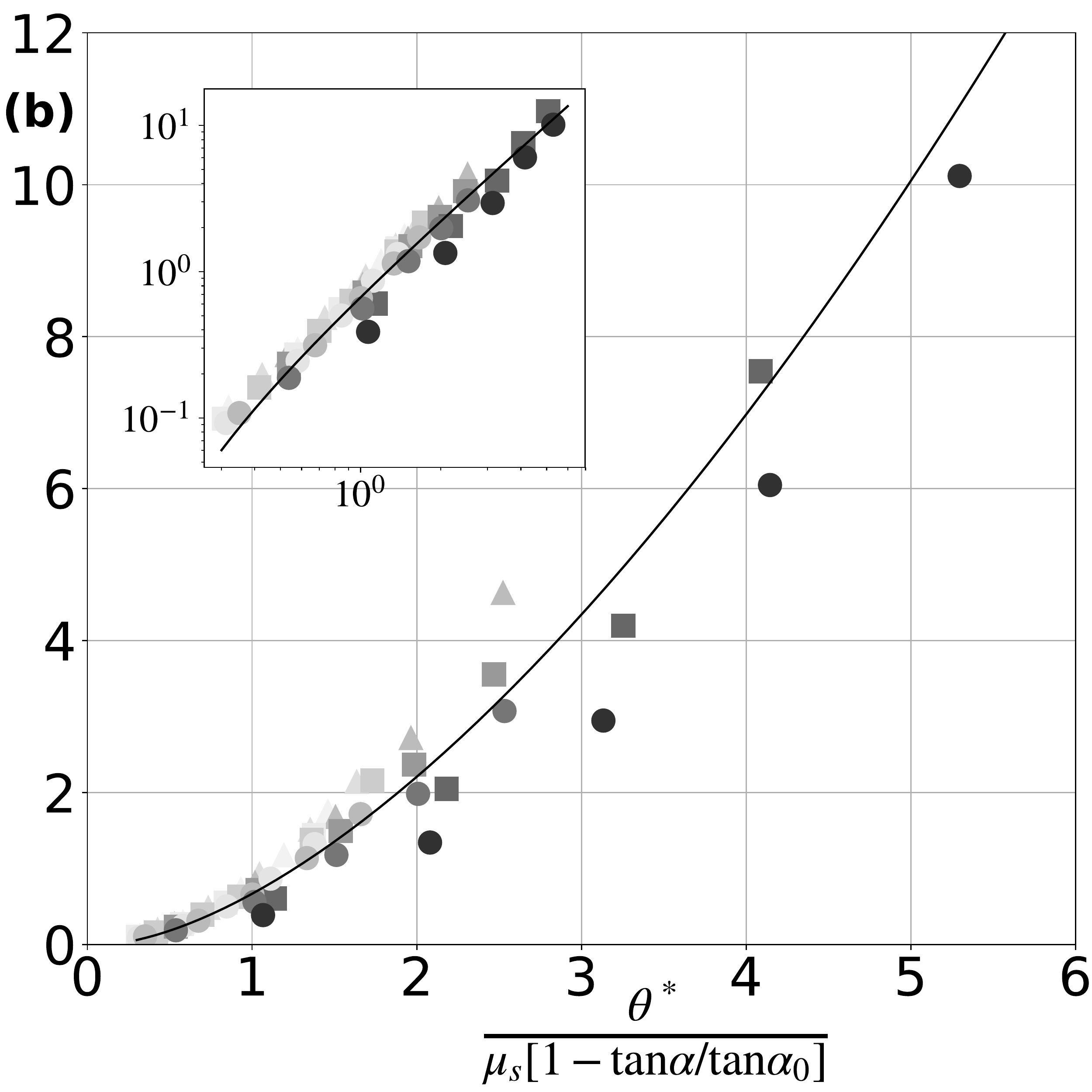}\\
\caption{\label{allData} Dimensionless sediment transport rate $Q_s^*$ as a function of the Shields number $\theta^*$ (a) and the rescaled Shields number (b) respectively, for variation of slope and specific density. The triangles, the squares and the circles denote simulations with specific density $\rho^p/\rho^f -1$ of $0.75$, $1.5$ and $3$ respectively and the darkness of the points is characteristic of the value of the slope angle $\alpha$ varied between 0.01 and 0.2 (1 and 12 degrees). The black line shows the power law best fit, reading $Q_s^* = a(\theta^*_m-0.1)^{b}$, with $a = 0.79$ and $b = 1.60$ and $\theta^*_m$ the modified Shields number.}
\end{figure}

Combining the variation of the slope and of the specific density, figure \ref{allData}a shows a complex picture of the phenomenon, with an important dispersion at given Shields number values resulting from the  coupling between the two mechanisms.
Evaluating the granular medium friction coefficient from dry inclined plane avalanche simulations ($\mu_s \simeq 0.4$), the results are plotted in figure \ref{allData}b as a function of the modified Shields number proposed in the previous section. All the data are shown to collapse on a single curve of power law close to 3/2, with some dispersion at high modified Shields number values. This collapse shows that the rescaling characterises at first order the effect of the slope and of the specific density variations on the sediment transport rate in gravity-driven turbulent bedload transport. While the small dispersion observed at high modified Shields number values highlights the limits of the proposed scaling law, its intuitive nature, its simplicity and the fact that it encompasses the different physical mechanisms represent a clear improvement compared with previous corrections.\\

\begin{figure}
\includegraphics[width=0.5\textwidth]{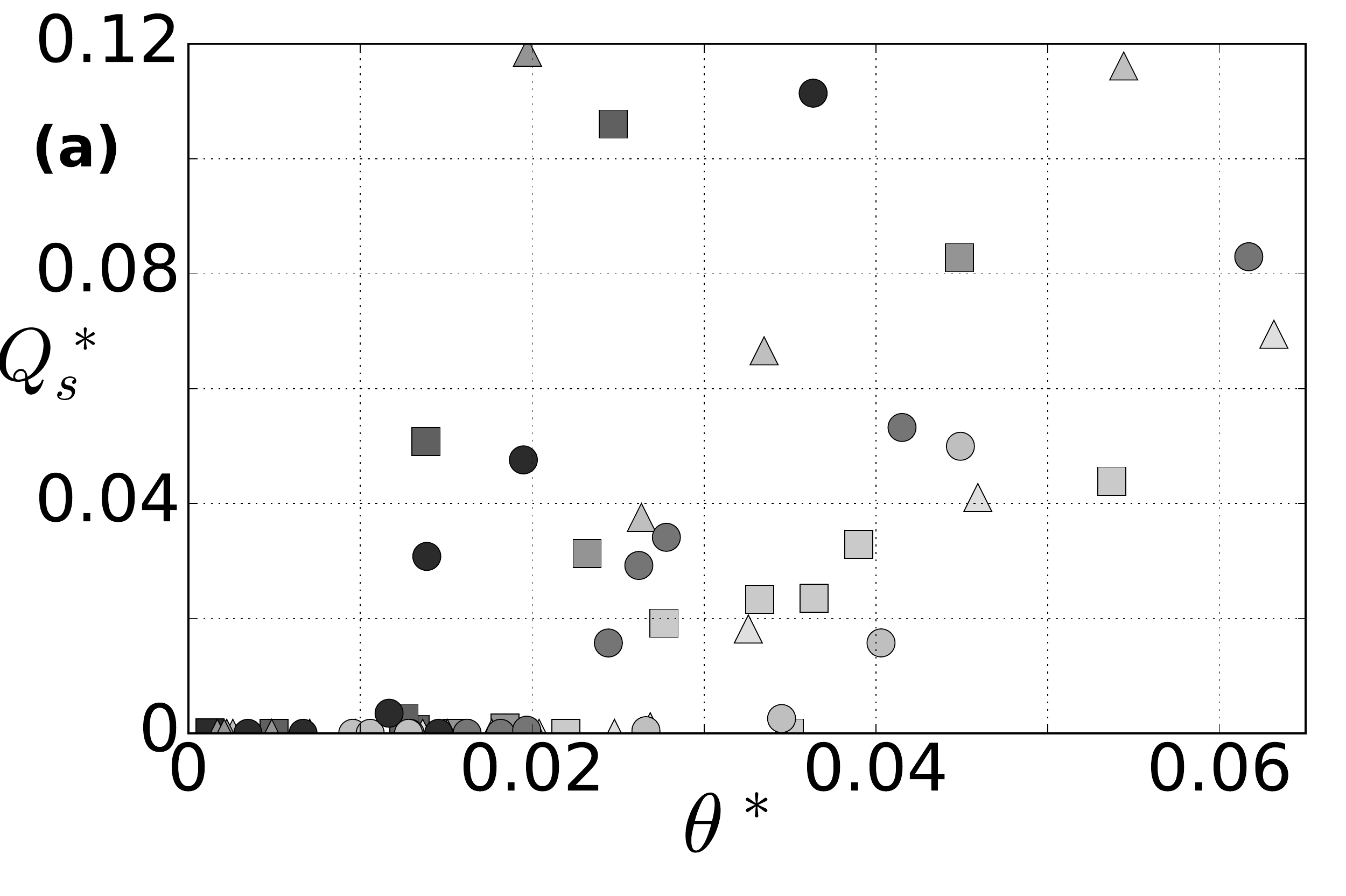}  \includegraphics[width=0.5\textwidth]{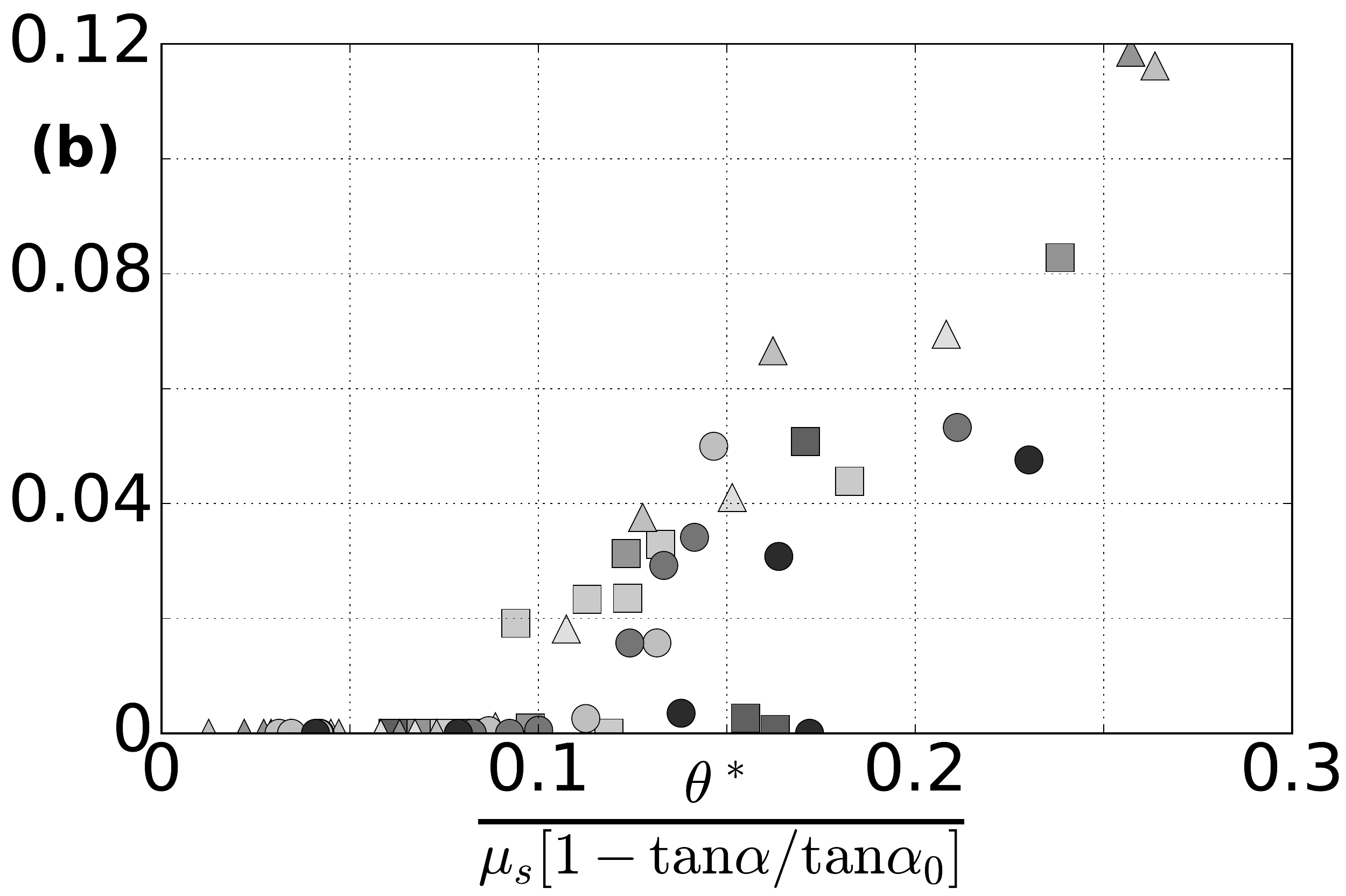}
\caption{\label{onset}Dimensionless sediment transport rate $Q_s^*$ as a function of the Shields number $\theta^*$ (a) and the rescaled Shields number (b) respectively, for variation of slope and specific density close to the onset of motion. The color code is the same as the previous figures. }
\end{figure}

Additional simulations have been performed close to the onset of motion to characterise the critical Shields number. The latter has been evaluated by forcing the flow slightly above the onset of motion before letting the simulations evolve at the given Shields number values in order to characterise the cessation threshold \citep{Ouriemi2007}. The results are shown in figure \ref{onset} and suggest that the proposed rescaling is valid at first order close to the critical Shields number. Even though the data exhibit more dispersion due to the complex characterisation of the onset of motion \citep{Clark2015}, the rather good collapse observed suggests that the proposed formulation allows us to define a unique critical Shields number independent of the slope and the specific density, which is close to $\theta^*_c/[\mu_s(1-\tan\alpha /\tan \alpha_0)] \simeq 0.1$ in the present simulations. \\

\begin{figure}
\includegraphics[width=\textwidth]{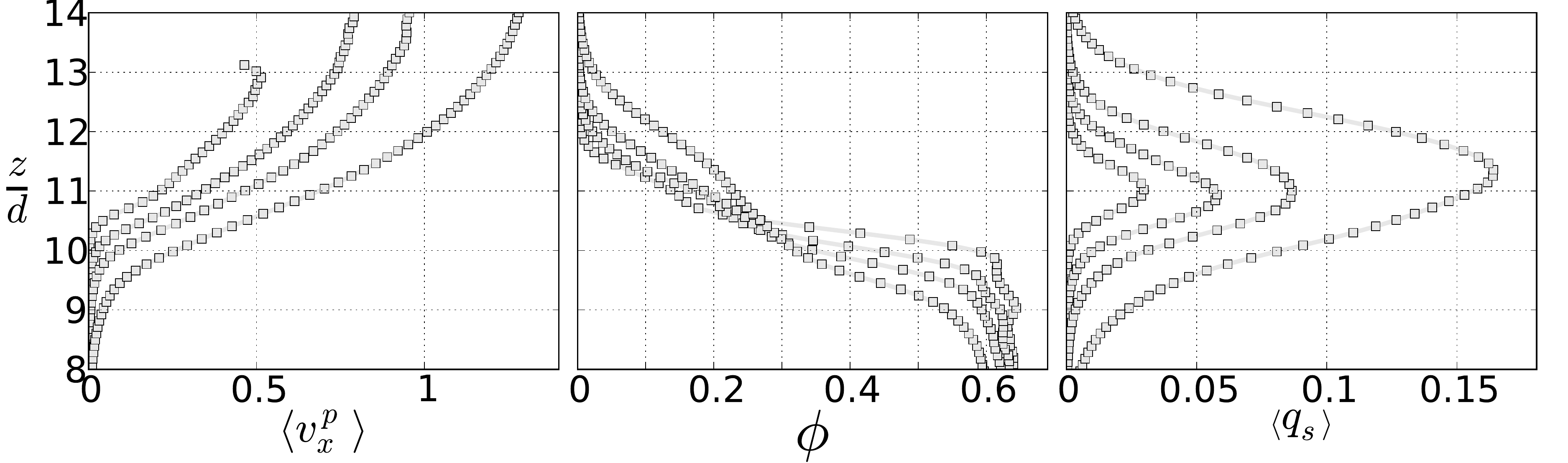} \\
\caption{\label{profiles1} Shields number influence on the solid velocity (m/s), volume fraction and transport rate density (m/s) depth profiles, at given slope ($\alpha = 0.02$, 1$^{\circ}$) and specific density ($\rho^p/\rho^f = 1.5$). The Shields number are respectively $\theta^* = 0.11$, $0.20$, $0.30$ and $0.49$ and correspond to the curves with increasing averaged velocity. }
\end{figure}

In order to illustrate the underlying mechanisms of the slope influence on the vertical granular flow structure, the solid depth profiles of volume fraction, $\phi$, average streamwise particle velocity, $<v_x^p>$ and transport rate density, $q_s = \phi <v_x^p>$ are shown in figure \ref{profiles1} for various Shields number values at constant $\tan \alpha/\tan \alpha_0 = 0.1$.
\begin{figure}
\includegraphics[width=\textwidth]{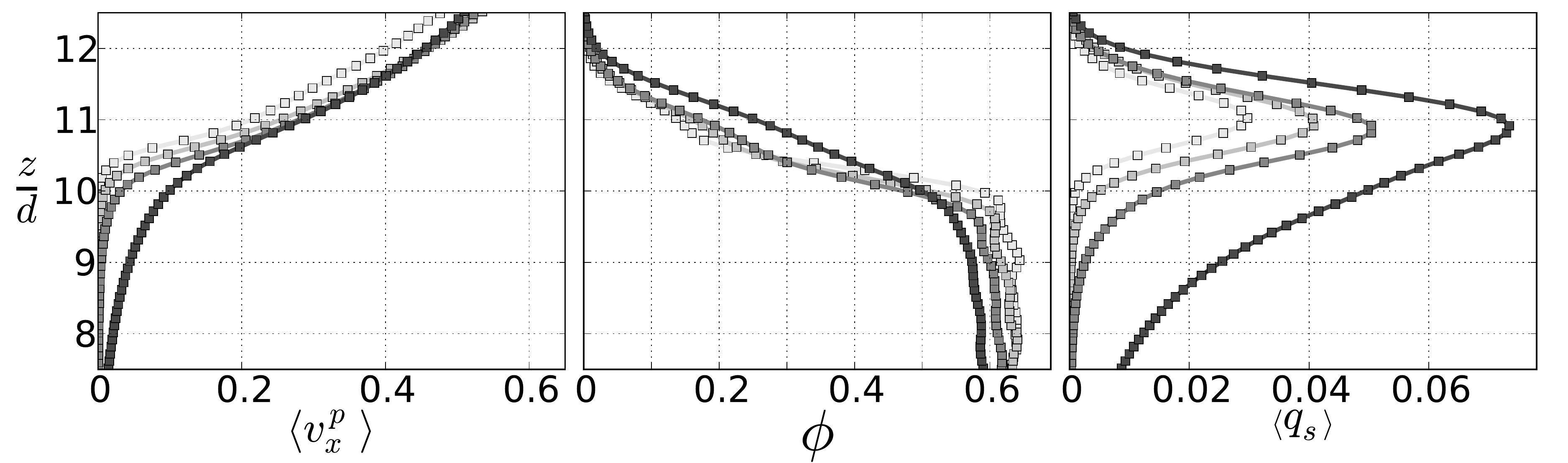} 
\caption{\label{profiles2} Slope influence on the solid velocity (m/s), volume fraction and transport rate density (m/s) depth profiles, at a given Shields number $\theta^* \sim 0.1$ and specific density ($\rho^p/\rho^f = 1.5$). The intensity of the points reflects the magnitude of the slope, which is varying between 1 and 8 degrees ($\alpha = {0.02,0.05,0.10,0.14}$).}
\end{figure}
The solid volume fraction depth profiles are characterised by the presence of a possible fixed point at the interface between the granular bed and the fluid flow, around $\phi = 0.3$. The Shields number increase leads to an increase of the solid volume fraction above the interface and a decrease below it, associated with an increase of the mobile granular layer thickness. The corresponding solid velocity profiles keep the same shape and exhibit a shift to higher values with increasing Shields number values, similarly to the transport rate density profiles which are broadening in the same time due to the increased mobile layer thickness.\\
The presence of the fixed point in the solid volume fraction profiles at a low relative slope inclination angle suggests that an appropriate non-dimensionalisation of the $z$ component could collapse all the solid volume fraction profiles obtained at different Shields number values. However, as we have seen in section \ref{GranularFlowStructure}, the vertical structure of the granular flow results from a competition between the slope influence on the lower part of the granular flow and the surface contribution from the dimensionless fluid bed shear stress, i.e. the Shields number. These two effects are mostly independent and they are dominant in the lower part and the upper part of the granular flow respectively. This is well observed when considering the variation of the slope influence ($\tan \alpha/\tan \alpha_0$) at a constant Shields number (see figure \ref{profiles2}). 

Increasing the slope induces an increasing vertical asymmetry in the transport rate profiles towards the bottom, resulting from a shift of the concentration shoulder to higher solid volume fraction and an increase of the solid velocity in the lower part of the mobile layer. The absence of a fixed point and this changing asymmetry in the solid volume fraction and transport rate density profiles show that the granular depth profiles cannot be self-similar in gravity-driven turbulent bedload transport. 

\section{Discussion}
\label{discussion}

The present theoretical analysis and numerical simulations of gravity-driven turbulent bedload transport demonstrate that the effect of the slope on the sediment transport rate is not restricted to the critical Shields number, but influence the whole transport rate formula, in agreement with part of the literature \citep{Smart1983,Smart1984,Rickenmann1991,Cheng2014}. It highlights the importance of the critical angle of debris flow in the transport rate prediction. The debris flow angle is representative of the effective resistance of the granular layer at rest and includes both the direct and indirect slope influence on the granular medium. Therefore, the analysis suggests that the key parameter in turbulent bedload transport is not the slope but the ratio between the tangent of the slope angle and the tangent of the critical angle of debris flow, $\tan \alpha/\tan \alpha_0$.\\

As a consequence, the specific density or density ratio plays a crucial role in the slope effect on the sediment transport rate. A variation of the density ratio has important consequences due to the associated modification of the critical debris flow angle and care should be taken when using model material in experiments. In particular, plastic particles are commonly used for sheet flows and bedload transport experiments \citep{Armanini2005,Larcher2007,Capart2011,Ni2015,RevilBaudard2015}, with density of approximately $\rho^p/ \rho^f = 1.2$, leading to a critical debris flow angle divided by four with respect to the classical density ratio for underwater natural sediment, $\rho^p/ \rho^f = 2.65$. This modification strongly influences both the sediment transport rate and the vertical structure of the granular flow, preventing a direct application of the results to natural sediment transport. In particular, the absence of concentration shoulder in the solid volume fraction profile measured by \citet{Capart2011} and \citet{Sumer1996} could be explained by the use of plastic particles in their experiments. Indeed, the resulting higher importance of the fluid contribution inside the granular layer tends to smooth out the concentration shoulder as seen in figure \ref{profiles2}.\\

Lastly, we discuss the common points and differences between the gravity-driven and the pressure-driven configurations in steady uniform turbulent bedload transport. To illustrate the comparison, we consider the entrainment mechanisms of the granular medium in the pressure-driven configuration, similarly to section \ref{GranularFlowStructure} for the gravity-driven case. Performing the same integration of the two-phase equation, the shear-to-normal stress ratio in the pressure-driven configuration reads (see appendix \ref{pressureDriven}):
\begin{equation}
\displaystyle \frac{\tau_{xz}^p(z)}{\tau_{zz}^p(z)}= \frac{\rho^p}{\rho^p-\rho^f} \tan \alpha + \frac{ \tau_b}{g \cos \alpha(\rho^p-\rho^f)\bar{\phi}_z (h_p - z)} +\frac{\partial P^f/\partial x}{ g \cos \alpha(\rho^p-\rho^f)\bar{\phi_z}}.
\label{stressRatioFinal_p}
\end{equation}
The two first terms on the right-hand side correspond to the direct influence of the slope on the granular phase and the Shields number dependence. They are exactly the same as for the gravity-driven case (see equation  \ref{stressRatioFinal}). The last term on the right-hand side represents the effect of the fluid flow induced by the pressure gradient inside the granular layer and does not vanish at zero slope.\\
Similarly to the gravity-driven configuration, increasing the slope at a constant Shields number decreases the resistance of the granular bed to entrainment and one might expect a similar consequence on the transport rate scaling. This indicates that the effect of the slope in pressure-driven configurations would not only be restricted to the critical Shields number as suggested in the literature \citep{Fernandez1976,Fredsoe1992,Chiew1994,Dey2003,Andreotti2013}, but also affects the whole transport rate formula, as evidenced by the experiments of \citet{Damgaard1997}. In addition, similarly to the gravity-driven case developed in this paper, the importance of the fluid flow inside the granular layer cannot be accounted for in the classical framework in term of Shields number  - i.e. dimensionless fluid shear stress at the \textit{top} of the granular bed - and should be considered when modelling pressure-driven bedload transport. This has already been achieved and verified experimentally in laminar bedload transport and leads to a transport rate scaling law in terms of dimensionless fluid flow rate \citep{Aussillous2013}.\\
Besides, the gravity-driven and pressure-driven configurations do not exhibit the same dependency on the slope. In the former, the fluid flow inside the granular bed is driven by the slope (eq. \ref{stressRatioFinal}), while in the latter it is driven by the pressure gradient and is independent of the slope (eq. \ref{stressRatioFinal_p}). As a consequence, the two configurations are different in nature and not equivalent in terms of slope dependency. Therefore, care should be taken to consider configuration where the subsurface fluid flow contribution is negligible when applying results established for pressure-driven flows (e.g. \citet{Chiew1994,Dey2003}) to field prediction and experimental gravity-driven configurations (e.g. \citet{Li1999,Wilcock2003,Karmaker2016} among others). 

\section{Conclusion}

Analysing turbulent bedload transport from a theoretical and numerical point of view, we attempted to clarify the mechanisms and origin of the slope influence and made a step toward a better understanding of the phenomenon. In particular, it has been evidenced that the classical modification of the critical Shields number relies on an expression of the buoyancy force which is not valid for uniform turbulent bedload transport. Focusing on gravity-driven configurations in steady uniform conditions, we have evidenced the entrainment mechanisms of the granular phase and shown the neglected importance of the fluid flow inside the bed. The relative importance of the latter with respect to the surface contribution characterised by the Shields number, affects the vertical structure of the granular flow and the sediment transport rate. The proposed modification of the Shields number to account macroscopically for these mechanisms has been shown to make all the present numerical data collapse onto a single master curve. It evidences that the key parameter in gravity-driven turbulent bedload transport is not the slope but the ratio between the tangent of the slope angle and the tangent of the critical debris flow angle. The difference is of importance when considering ideal particles for which the density (e.g. plastic) and the shape (e.g. spheres) are different from natural sediments. In addition, the theoretical analysis has evidenced a difference in nature between gravity-driven and pressure-driven configurations with respect to the slope influence. This result underlines the necessity to differentiate the two configurations in the analysis of slope influence on turbulent bedload transport. \\
The present contribution provides a better understanding of the slope influence in idealised configurations and represents a step in the understanding and prediction of the slope influence in turbulent bedload transport. In order to go further, it would be interesting to perform precisely controlled experiments and fully resolved simulations at varying slope inclination angle, to validate the present approach and characterise the importance of the neglected mechanisms such as the submergence, the lift force and the coherent structures.

\section*{Acknowledgement}

We would like to thank the three anonymous reviewers for their interesting and constructive questions. We are grateful to Ashley Dudill for English corrections. RM would like to thank Thomas P\"ahtz, Alain Recking and Laurent Lacaze for fruitful discussions.
This research was supported by Irstea (formerly Cemagref), the French national research agency project SegSed ANR-16-CE01-0005, the labex OSUG@2020 and the French Institut National des Sciences de l’Univers
program EC2CO-BIOHEFECT and EC2CO-LEFE MODSED. 

\appendix

\section{Pressure-driven vertical granular flow structure}
\label{pressureDriven}
For the pressure-driven case, only the streamwise fluid momentum balance is modified, by replacing the driving slope term with a pressure gradient:
\begin{equation}
0 = \frac{\partial S_{xz}^f}{\partial z} + \frac{\partial R_{xz}^f}{\partial z} - (1-\phi) \frac{\partial P^f}{\partial x} - n \left<{f_f^p}_x\right>^p. 
\label{fluidXZ_p}
\end{equation}
Contrary to the gravity-driven case, there exists a pressure gradient in the streamwise direction so that the buoyancy force has a streamwise component which can be expressed as $-\phi \frac{\partial P^f}{\partial x}$ and leads to the following reformulation: 
\begin{equation}
0 = \frac{\partial S_{xz}^f}{\partial z} + \frac{\partial R_{xz}^f}{\partial z} - \frac{\partial P^f}{\partial x} - n \left<{f_f^p}_x\right>^p. 
\label{fluidXZ_pBIS}
\end{equation}
Performing the same manipulation of the equations as for the gravity driven case, the shear-to-normal granular stress ratio can be shown to read: 
\begin{multline}
\displaystyle  \frac{\tau_{xz}^p(z)}{\tau_{zz}^p(z)}= \frac{\rho^f g \sin \alpha  \displaystyle \displaystyle \int_{z}^{h_p}{\phi(\zeta) d \zeta}}{g \cos \alpha(\rho^p-\rho^f)\displaystyle \int_{z}^{h_p}{\phi(\zeta) d \zeta}} + \frac{ \tau_b + \frac{\partial P^f}{\partial x}\displaystyle \int_{z}^{h_p}{d \zeta}}{g \cos \alpha(\rho^p-\rho^f)\displaystyle \int_{z}^{h_p}{\phi(\zeta) d \zeta}},
\end{multline}
leading to the final equation:
\begin{equation}
\displaystyle \frac{\tau_{xz}^p(z)}{\tau_{zz}^p(z)}= \frac{\rho^p}{\rho^p-\rho^f} \tan \alpha + \frac{ \tau_b}{g \cos \alpha(\rho^p-\rho^f)\bar{\phi}_z (h_p - z)} +\frac{\partial P^f/\partial x}{ g \cos \alpha(\rho^p-\rho^f)\bar{\phi_z}}.
\label{stressRatioFinal_pA}
\end{equation}\\

\definecolor{grey0}{gray}{ 0.9475 }
\definecolor{grey1}{gray}{ 0.9475 }
\definecolor{grey2}{gray}{ 0.9475 }
\definecolor{grey3}{gray}{ 0.9475 }
\definecolor{grey4}{gray}{ 0.9475 }
\definecolor{grey5}{gray}{ 0.86875 }
\definecolor{grey6}{gray}{ 0.86875 }
\definecolor{grey7}{gray}{ 0.86875 }
\definecolor{grey8}{gray}{ 0.86875 }
\definecolor{grey9}{gray}{ 0.86875 }
\definecolor{grey10}{gray}{ 0.7375 }
\definecolor{grey11}{gray}{ 0.7375 }
\definecolor{grey12}{gray}{ 0.7375 }
\definecolor{grey13}{gray}{ 0.7375 }
\definecolor{grey14}{gray}{ 0.7375 }
\definecolor{grey15}{gray}{ 0.920416666667 }
\definecolor{grey16}{gray}{ 0.920416666667 }
\definecolor{grey17}{gray}{ 0.920416666667 }
\definecolor{grey18}{gray}{ 0.920416666667 }
\definecolor{grey19}{gray}{ 0.801041666667 }
\definecolor{grey20}{gray}{ 0.801041666667 }
\definecolor{grey21}{gray}{ 0.801041666667 }
\definecolor{grey22}{gray}{ 0.801041666667 }
\definecolor{grey23}{gray}{ 0.801041666667 }
\definecolor{grey24}{gray}{ 0.602083333333 }
\definecolor{grey25}{gray}{ 0.602083333333 }
\definecolor{grey26}{gray}{ 0.602083333333 }
\definecolor{grey27}{gray}{ 0.602083333333 }
\definecolor{grey28}{gray}{ 0.602083333333 }
\definecolor{grey29}{gray}{ 0.403125 }
\definecolor{grey30}{gray}{ 0.403125 }
\definecolor{grey31}{gray}{ 0.403125 }
\definecolor{grey32}{gray}{ 0.403125 }
\definecolor{grey33}{gray}{ 0.403125 }
\definecolor{grey34}{gray}{ 0.894583333333 }
\definecolor{grey35}{gray}{ 0.894583333333 }
\definecolor{grey36}{gray}{ 0.894583333333 }
\definecolor{grey37}{gray}{ 0.894583333333 }
\definecolor{grey38}{gray}{ 0.894583333333 }
\definecolor{grey39}{gray}{ 0.729166666667 }
\definecolor{grey40}{gray}{ 0.729166666667 }
\definecolor{grey41}{gray}{ 0.729166666667 }
\definecolor{grey42}{gray}{ 0.729166666667 }
\definecolor{grey43}{gray}{ 0.729166666667 }
\definecolor{grey44}{gray}{ 0.4625 }
\definecolor{grey45}{gray}{ 0.4625 }
\definecolor{grey46}{gray}{ 0.4625 }
\definecolor{grey47}{gray}{ 0.4625 }
\definecolor{grey48}{gray}{ 0.4625 }
\definecolor{grey49}{gray}{ 0.191666666667 }
\definecolor{grey50}{gray}{ 0.191666666667 }
\definecolor{grey51}{gray}{ 0.191666666667 }
\definecolor{grey52}{gray}{ 0.191666666667 }
\definecolor{grey53}{gray}{ 0.191666666667 }

\section{Detailed characteristics of the numerical runs}
\label{runCharacteristics}

Table \ref{tableParam} presents the detailed characteristics  and the associated symbols of the simulations performed far from the threshold of motion, excluding the simulations of figure \ref{onset} which are secondary. The Reynolds number, $Re = U h/\nu^f$, the Froude number  $Fr = U/\sqrt{g_z h}$ and the relative submergence $h/d$ are evaluated considering the mean fluid velocity $U$ inside water depth $h$, defined as the difference between the position of the water free surface and the maximum of the Reynolds stresses. This is consistent with the definition adopted for the Shields number, based on the maximum Reynolds stresses and equivalent to $\theta^* = h \sin \alpha/[(\rho^p/\rho^f -1)d]$.
\begin{table}
 \begin{center}
  \begin{tabular}{ccccccc}
 $\alpha$ &  \hspace{0.5cm} $\rho^p/\rho^f - 1$ \hspace{0.5cm}  & \hspace{0.5cm} $\theta^*$ \hspace{0.5cm}& $Re$  & \hspace{0.5cm}  $Fr$  \hspace{0.5cm} & \hspace{0.5cm}  $h/d$  \hspace{0.5cm}  &  symbol\\
\noalign{\smallskip}\hline\noalign{\smallskip}
0.0126 	& 0.75 	& 0.11 	& 35000 	& 1.21 	& 7.3 	& $\color{grey0}{ \blacktriangle }$ \\
0.0126 	& 0.75 	& 0.21 	& 84000 	& 1.24 	& 13.0 	& $\color{grey1}{ \blacktriangle }$ \\
0.0126 	& 0.75 	& 0.34 	& 175000 	& 1.28 	& 20.7 	& $\color{grey2}{ \blacktriangle }$ \\
0.0126 	& 0.75 	& 0.43 	& 257000 	& 1.31 	& 26.3 	& $\color{grey3}{ \blacktriangle }$ \\
0.0126 	& 0.75 	& 0.52 	& 356000 	& 1.35 	& 32.1 	& $\color{grey4}{ \blacktriangle }$ \\
0.0315 	& 0.75 	& 0.13 	& 16000 	& 1.66 	& 3.5 	& $\color{grey5}{ \blacktriangle }$ \\
0.0315 	& 0.75 	& 0.22 	& 34000 	& 1.68 	& 5.8 	& $\color{grey6}{ \blacktriangle }$ \\
0.0315 	& 0.75 	& 0.31 	& 57000 	& 1.75 	& 8.0 	& $\color{grey7}{ \blacktriangle }$ \\
0.0315 	& 0.75 	& 0.4 	& 85000 	& 1.8 	& 10.2 	& $\color{grey8}{ \blacktriangle }$ \\
0.0315 	& 0.75 	& 0.49 	& 116000 	& 1.85 	& 12.3 	& $\color{grey9}{ \blacktriangle }$ \\
0.063 	& 0.75 	& 0.1 	& 6000 	& 2.01 	& 1.7 	& $\color{grey10}{ \blacktriangle }$ \\
0.063 	& 0.75 	& 0.2 	& 15000 	& 2.13 	& 2.8 	& $\color{grey11}{ \blacktriangle }$ \\
0.063 	& 0.75 	& 0.3 	& 26000 	& 2.24 	& 4.0 	& $\color{grey12}{ \blacktriangle }$ \\
0.063 	& 0.75 	& 0.39 	& 40000 	& 2.32 	& 5.2 	& $\color{grey13}{ \blacktriangle }$ \\
0.063 	& 0.75 	& 0.5 	& 64000 	& 2.36 	& 7.0 	& $\color{grey14}{ \blacktriangle }$ \\
0.0191 	& 1.5 	& 0.11 	& 62000 	& 1.53 	& 9.2 	& $\color{grey15}{ \blacksquare }$ \\
0.0191 	& 1.5 	& 0.21 	& 156000 	& 1.55 	& 16.9 	& $\color{grey16}{ \blacksquare }$ \\
0.0191 	& 1.5 	& 0.3 	& 279000 	& 1.58 	& 24.5 	& $\color{grey17}{ \blacksquare }$ \\
0.0191 	& 1.5 	& 0.49 	& 605000 	& 1.66 	& 39.8 	& $\color{grey18}{ \blacksquare }$ \\
0.04775 	& 1.5 	& 0.12 	& 28000 	& 2.08 	& 4.4 	& $\color{grey19}{ \blacksquare }$ \\
0.04775 	& 1.5 	& 0.21 	& 58000 	& 2.11 	& 7.1 	& $\color{grey20}{ \blacksquare }$ \\
0.04775 	& 1.5 	& 0.28 	& 94000 	& 2.09 	& 9.8 	& $\color{grey21}{ \blacksquare }$ \\
0.04775 	& 1.5 	& 0.41 	& 159000 	& 2.23 	& 13.4 	& $\color{grey22}{ \blacksquare }$ \\
0.04775 	& 1.5 	& 0.52 	& 234000 	& 2.3 	& 17.0 	& $\color{grey23}{ \blacksquare }$ \\
0.0955 	& 1.5 	& 0.11 	& 11000 	& 2.55 	& 2.1 	& $\color{grey24}{ \blacksquare }$ \\
0.0955 	& 1.5 	& 0.2 	& 27000 	& 2.65 	& 3.7 	& $\color{grey25}{ \blacksquare }$ \\
0.0955 	& 1.5 	& 0.3 	& 49000 	& 2.76 	& 5.3 	& $\color{grey26}{ \blacksquare }$ \\
0.0955 	& 1.5 	& 0.39 	& 72000 	& 2.85 	& 6.7 	& $\color{grey27}{ \blacksquare }$ \\
0.0955 	& 1.5 	& 0.49 	& 102000 	& 2.98 	& 8.2 	& $\color{grey28}{ \blacksquare }$ \\
0.14325 	& 1.5 	& 0.11 	& 8000 	& 2.71 	& 1.6 	& $\color{grey29}{ \blacksquare }$ \\
0.14325 	& 1.5 	& 0.21 	& 19000 	& 3.02 	& 2.6 	& $\color{grey30}{ \blacksquare }$ \\
0.14325 	& 1.5 	& 0.31 	& 36000 	& 3.37 	& 3.8 	& $\color{grey31}{ \blacksquare }$ \\
0.14325 	& 1.5 	& 0.39 	& 53000 	& 3.72 	& 4.6 	& $\color{grey32}{ \blacksquare }$ \\
0.14325 	& 1.5 	& 0.48 	& 79000 	& 4.1 	& 5.6 	& $\color{grey33}{ \blacksquare }$ \\
0.0253 	& 3.0 	& 0.11 	& 139000 	& 1.83 	& 14.0 	& $\color{grey34}{ \bullet }$ \\
0.0253 	& 3.0 	& 0.21 	& 346000 	& 1.85 	& 25.5 	& $\color{grey35}{ \bullet }$ \\
0.0253 	& 3.0 	& 0.31 	& 615000 	& 1.87 	& 37.1 	& $\color{grey36}{ \bullet }$ \\
0.0253 	& 3.0 	& 0.4 	& 941000 	& 1.91 	& 48.6 	& $\color{grey37}{ \bullet }$ \\
0.0253 	& 3.0 	& 0.5 	& 1324000 	& 1.95 	& 60.2 	& $\color{grey38}{ \bullet }$ \\
0.065 	& 3.0 	& 0.1 	& 44000 	& 2.5 	& 5.3 	& $\color{grey39}{ \bullet }$ \\
0.065 	& 3.0 	& 0.2 	& 113000 	& 2.52 	& 9.8 	& $\color{grey40}{ \bullet }$ \\
0.065 	& 3.0 	& 0.3 	& 204000 	& 2.55 	& 14.4 	& $\color{grey41}{ \bullet }$ \\
0.065 	& 3.0 	& 0.4 	& 316000 	& 2.62 	& 19.0 	& $\color{grey42}{ \bullet }$ \\
0.065 	& 3.0 	& 0.5 	& 448000 	& 2.68 	& 23.6 	& $\color{grey43}{ \bullet }$ \\
0.129 	& 3.0 	& 0.11 	& 22000 	& 3.08 	& 2.9 	& $\color{grey44}{ \bullet }$ \\
0.129 	& 3.0 	& 0.2 	& 53000 	& 3.13 	& 5.2 	& $\color{grey45}{ \bullet }$ \\
0.129 	& 3.0 	& 0.3 	& 96000 	& 3.22 	& 7.5 	& $\color{grey46}{ \bullet }$ \\
0.129 	& 3.0 	& 0.4 	& 149000 	& 3.32 	& 9.9 	& $\color{grey47}{ \bullet }$ \\
0.129 	& 3.0 	& 0.5 	& 216000 	& 3.45 	& 12.3 	& $\color{grey48}{ \bullet }$ \\
0.194 	& 3.0 	& 0.1 	& 13000 	& 3.26 	& 2.0 	& $\color{grey49}{ \bullet }$ \\
0.194 	& 3.0 	& 0.2 	& 32000 	& 3.37 	& 3.5 	& $\color{grey50}{ \bullet }$ \\
0.194 	& 3.0 	& 0.3 	& 61000 	& 3.73 	& 5.1 	& $\color{grey51}{ \bullet }$ \\
0.194 	& 3.0 	& 0.39 	& 101000 	& 4.15 	& 6.6 	& $\color{grey52}{ \bullet }$ \\
0.194 	& 3.0 	& 0.5 	& 156000 	& 4.55 	& 8.3 	& $\color{grey53}{ \bullet }$ \\
  \end{tabular}
  \caption{Parameters of the simulations studied and symbol correspondence.}\label{tableParam}
 \end{center}
\end{table}

\bibliography{NewBibliographyBIS}
\bibliographystyle{jfm}

\end{document}